\documentclass[twocolumn]{aastex63}

\graphicspath{{./}{figures/}}
\usepackage{pifont}
\usepackage{multirow}
\usepackage{amsmath}

\begin{document}

\title{Kilonova and Optical Afterglow from Binary Neutron Star Mergers. II. Optimal Search Strategy for Serendipitous Observations and Target-of-opportunity Observations of Gravitational-wave Triggers}

\author[0000-0002-9195-4904]{Jin-Ping Zhu}
\affil{Department of Astronomy, School of Physics, Peking University, Beijing 100871, China; \url{zhujp@pku.edu.cn}}

\author[0000-0002-9188-5435]{Shichao Wu}
\affiliation{Max-Planck-Institut f{\"u}r Gravitationsphysik (Albert-Einstein-Institut), D-30167 Hannover, Germany; \url{shichao.wu@aei.mpg.de}}
\affiliation{Leibniz Universit{\"a}t Hannover, D-30167 Hannover, Germany}

\author[0000-0001-6374-8313]{Yuan-Pei Yang}
\affiliation{South-Western Institute for Astronomy Research, Yunnan University, Kunming, Yunnan, People’s Republic of China; \url{ypyang@ynu.edu.cn}}

\author[0000-0001-7649-6792]{Chang Liu}
\affiliation{Department of Astronomy, School of Physics, Peking University, Beijing 100871, China; \url{zhujp@pku.edu.cn}}
\affiliation{Kavli Institute for Astronomy and Astrophysics, Peking University, Beijing 100871, China}

\author[0000-0002-9725-2524]{Bing Zhang}
\affiliation{Nevada Center for Astrophysics, University of Nevada, Las Vegas, NV 89154, USA; \url{bing.zhang@unlv.edu}}
\affiliation{Department of Physics and Astronomy, University of Nevada, Las Vegas, NV 89154, USA}

\author[0000-0002-7348-4304]{Hao-Ran Song}
\affiliation{Department of Astronomy, Beijing Normal University, Beijing 100875, China}

\author[0000-0002-3100-6558]{He Gao}
\affiliation{Department of Astronomy, Beijing Normal University, Beijing 100875, China}

\author[0000-0002-1932-7295]{Zhoujian Cao}
\affiliation{Department of Astronomy, Beijing Normal University, Beijing 100875, China}

\author[0000-0002-1067-1911]{Yun-Wei Yu}
\affiliation{Institute of Astrophysics, Central China Normal University, Wuhan 430079, China}

\author[0000-0001-7402-4927]{Yacheng Kang}
\affiliation{Department of Astronomy, School of Physics, Peking University, Beijing 100871, China; \url{zhujp@pku.edu.cn}}
\affiliation{Kavli Institute for Astronomy and Astrophysics, Peking University, Beijing 100871, China}

\author[0000-0002-1334-8853]{Lijing Shao}
\affiliation{Kavli Institute for Astronomy and Astrophysics, Peking University, Beijing 100871, China}
\affiliation{National Astronomical Observatories, Chinese Academy of Sciences, Beijing 100012, China}

\begin{abstract}

In the second work of this series, we explore the optimal search strategy for serendipitous and gravitational-wave-triggered target-of-opportunity (ToO) observations of kilonovae and optical short-duration gamma-ray burst (sGRB) afterglows from binary neutron star (BNS) mergers, assuming that cosmological kilonovae are AT2017gfo-like (but with viewing-angle dependence) and that the properties of afterglows are consistent with those of cosmological sGRB afterglows. A one-day cadence serendipitous search strategy with an exposure time of $\sim30\,$s can always achieve an optimal search strategy of kilonovae and afterglows for various survey projects. We show that the optimal detection rates of the kilonovae (afterglows) are $\sim0.3/0.6/1/20\,$yr$^{-1}$ ({$\sim50/60/100/800\,$yr$^{-1}$}) for ZTF/Mephisto/WFST/LSST, respectively. A better search strategy for SiTian than the current design is to increase the exposure time. In principle, a fully built SiTian can detect $\sim7({2000})\,$yr$^{-1}$ kilonovae (afterglows). {{Population properties of electromagnetic (EM) signals detected via the serendipitous observations are studied in detail. For ToO observations, we predict that one can detect $\sim11\,{\rm{yr}}^{-1}$ BNS gravitational wave (GW) events during the fourth observing run (O4) by considering an exact duty cycle of the third observing run. The median GW sky localization area is expected to be $\sim10\,{\rm{deg}}^2$ for detectable BNS GW events. For O4, we predict that ZTF/Mephisto/WFST/LSST can detect $\sim5/4/3/3$ kilonovae ($\sim1/1/1/1$ afterglows) per year, respectively. The GW detection rates, GW population properties, GW sky localizations, and optimistic ToO detection rates of detectable EM counterparts for BNS GW events at the Advanced Plus, LIGO Voyager and ET\&CE eras are detailedly simulated in this paper.}}

\end{abstract}


\section{Introduction} \label{sec:intro}

Kilonovae \citep{li1998,metzger2010} and short-duration gamma-ray bursts \citep[sGRB;][]{pacynski1986,paczynski1991,eichler1989,narayan1992,zhangb2018} have long been thought to originate from binary neutron star (BNS) and neutron star--black hole (NSBH) mergers. The interaction of the sGRB relativistic jets with the surrounding interstellar medium would produce bright afterglow emissions from X-ray to radio\footnote{If BNS and NSBH mergers occur in active galactic nucleus \citep[e.g.,][]{cheng1999,mckernan2020} accretion disks, sGRB relativistic jets would always be choked and kilonova emissions would be outshone by the disk emission \citep{zhu2021neutron,perna2021}. The choked jets and subsequent jet-cocoon and ejecta shock breakouts can generate high-energy neutrinos which may significantly contribute diffuse neutrino background \citep{zhu2021highenergy2,zhu2021highenergy1}.} \citep{rees1992,meszaros1993,paczynski1993,meszaros1997,sari1998,gao2013NAR}.

On 2017 August 17, the first BNS gravitational wave (GW) event, i.e., GW170817, was detected by the Advanced Laser Interferometer Gravitational Wave Observatory \citep[LIGO;][]{harry2010,aasi2015} and the Advanced Virgo \citep{acernese2015} detectors \citep{abbott2017gw170817}. This BNS GW event has been subsequently confirmed in connection with an sGRB \citep[GRB170817A;][]{abbott2017gravitational,goldstein2017,savchenko2017,zhangbb2018}, an ultraviolet–optical–infrared kilonova \citep[AT2017gfo;][]{abbott2017multimessenger,andreoni2017,arcavi2017,chornock2017,coulter2017,covino2017,cowperthwaite2017,diaz2017,drout2017,evans2017,hu2017,kasliwal2017,kilpatrick2017,lipunov2017,mccully2017,nicholl2017,pian2017,shappee2017,smartt2017,soaressantos2017,tanvir2017,utsumi2017,valenti2017,villar2017} and a broadband off-axis jet afterglow \citep{alexander2017,haggard2017,hallinan2017,margutti2017,troja2017,troja2018,troja2020,davanzo2018,dobie2018,lazzati2018,lyman2018,xie2018,piro2019,ghirlanda2019}. The multi-messenger observations of this BNS merger provided smoking-gun evidence for the long-hypothesized origin of sGRBs and kilonovae, and heralded the advent of the GW-led astronomy era. 

To date, except for AT2017gfo, other kilonova candidates were all detected in superposition with decaying sGRB afterglows \citep[e.g.,][]{berger2013,tanvir2013,fan2013,gao2015,gao2017,jin2015,jin2016,jin2020,yang2015,gompertz2018,ascenzi2019,rossi2020,ma2021,fong2021,wu2021,yuan2021}. {{Interestingly, a bright kilonova candidate was found to be associated with a long-duration GRB\,211211A in recent \citep[e.g.,][]{rastinejad2022,troja2022,yang2022lGRB,zhu2022lGRB}.}} One possible reason almost all kilonova candidates were detected in GRB afterglows is that most BNS and NSBH mergers are far away from us. Their associated kilonova signals may be too faint to be directly detected by present survey projects. However, thanks to the beaming effect of relativistic jets, in\defcitealias{zhu2021kilonovaafterglow}{Paper I}\citetalias{zhu2021kilonovaafterglow} of this series \citep{zhu2021kilonovaafterglow}, we have shown that a large fraction of cosmological afterglows could be much brighter than the associated kilonovae if the jets move towards or close to the line of sight. Bright afterglow emissions would help us detect potential associated kilonova emissions. On the other hand, a too bright afterglow would also affect on the detectability of the associated kilonova.

Catching more kilonovae and afterglows by current and future survey projects would be helpful for expanding our knowledge about the population properties of these events. \cite{kasliwal2020,mohite2021} constrained the population properties of kilonovae based on the non-detection of GW-triggered follow-up observations during O3. Although the properties of kilonova and afterglow emissions from BNS and NSBH mergers can be reasonably well predicted, their low luminosities and fast evolution nature compared with supernova emission make it difficult to detect them using the traditional time-domain survey projects. Several works in the literature have studied the detection rates and search strategy for kilonovae by serendipitous observations \citep[e.g.,][]{metzger2012,coughlin2017,coughlin2020implications,rosswog2017,scolnic2018,setzer2019,saguescarracedo2021,zhu2021neutron,andreoni2021,almualla2021,chase2021}. Because afterglow emission could significantly affect the observation of a fraction of kilonova events, one cannot ignore the effect of afterglow emission when considering the search strategy and detectability of kilonova emission. In the second work of this series, we will perform a detailed study on optimizing serendipitous detections of both kilonovae and optical afterglows with different cadences, filters, and exposure times for several present and future survey projects. The survey projects we consider in this work include the Zwicky Transient Facility \citep[ZTF;][]{bellm2019,masci2019}, the Multi-channel Photometric Survey Telescope\footnote{\url{http://www.mephisto.ynu.edu.cn/site/}} (Mephisto; Er et al. 2022, in preparation), the Wide Field Survey Telescope (WFST; et al. Kong et al. 2022, in preparation), the Large Synoptic Survey Telescope \citep[LSST;][]{lsst2009}, and the SiTian Projects \citep[SiTian;][]{liu2021}. We note that (1) NSBH mergers may have a lower event rate density; (2) NSBH kilonovae may be dimmer than BNS kilonovae \citep[e.g.,][]{zhu2020}; (3) most NSBH mergers in the universe are likely plunging events \citep[e.g.,][]{abbott2021observation,zappa2019,drozda2020,zhu2021population,zhu2021no,broekgaarden2021,hu2022}. As a result, the detection rates of kilonova and afterglow emissions from NSBH mergers should be much lower than those from BNS mergers \citep{zhu2021kilonova}. In the following calculations, we only consider sGRB, kilonova and afterglow emissions from BNS mergers.

Furthermore, with the upgrade and iteration of GW observatories, numerous BNS mergers from the distant universe will be discovered. Future foreseeable GW observations will give a better constraint on the localization for a fraction of BNS GW events, which will benefit the search for associated electromagnetic (EM) counterparts. For example, some GW sources will be localized to $\sim10\,{\rm deg}^2$ by the network including the Advanced LIGO, Advanced Virgo, and KAGRA GW detectors \citep{abbott2020prospects,frostig2021}. Therefore, taking advantage of target-of-opportunity (ToO) follow-up observations of GW triggers will greatly improve the search efficiency of kilonovae and afterglows, {{although \cite{petrov2022} recently suggested that the previous expectations for the GW sky localization may be too optimistic}}. The kilonova follow-up campaigns by specific survey projects, e.g., ZTF, LSST, and the Wide-Field Infrared Transient Explorer, for GW BNS mergers in the near GW era have been simulated recently \citep{saguescarracedo2021,cowperthwaite2019,frostig2021}. In this paper, we present detailed calculations of the BNS detectability by the GW detectors in the next $15\,$yr and the associated EM detectability for GW-triggered ToO observations. 

The paper is organized as follows. The physical models are briefly presented in Section \ref{sec:model}. More details of our models have been presented in \citetalias{zhu2021kilonovaafterglow}. The search strategy and detectability of kilonova and afterglow emissions for time-domain survey observations are studied in Section \ref{sec:serendiption}. We also perform some calculations for the EM detection rates by some specific survey projects. In Section \ref{sec:too}, we simulate the GW detection and subsequent detectability of EM ToO follow up observations for networks of 2nd-, 2.5th-, and 3rd-generation GW detectors. Finally, we summarize our conclusions and present some discussions in Section \ref{sec:conclusion}. A standard $\Lambda$CDM cosmology with $H_0 = 67.8\,{\rm km}\,{\rm s}^{-1}\,{\rm Mpc}^{-1}$, $\Omega_\Lambda = 0.692$, and $\Omega_{\rm m} = 0.308$ \citep{planck2016} is applied in this paper.

\section{Modelling \label{sec:model}}

\subsection{{{Redshift Distribution and EM Properties of Simulated BNS Populations}}}

The total number of BNS mergers in the universe can be estimated as \citep[e.g.,][]{sun2015}

\begin{equation}
\label{equ:merger_rate}
    \dot{N}_{\rm BNS} \approx \int_{0}^{z_{\rm max}}\frac{\dot{\rho}_{0,{\rm BNS}}f(z)}{1 + z}\frac{dV(z)}{dz}dz,
\end{equation}
where $\dot{\rho}_{0,{\rm BNS}}$ is the local BNS event rate density, $f(z)$ is the dimensionless redshift distribution factor, and $z_{\rm max}$ is the maximum redshift for BNS mergers. The comoving volume element $dV(z)/dz$ in Equation (\ref{equ:merger_rate}) is

\begin{equation}
    \frac{dV}{dz} = \frac{c}{H_0}\frac{4\pi D_{\rm L}^2}{(1 + z)^2 \sqrt{\Omega_\Lambda + \Omega_{\rm m}(1 + z)^3}},
\end{equation}
where $c$ is the speed of light and $D_{\rm L}$ is the luminosity distance, which is expressed as

\begin{equation}
    D_{\rm L} = (1 + z) \frac{c}{H_0} \int_0^z \frac{dz}{\sqrt{\Omega_\Lambda + \Omega_{\rm m}(1 + z)^3}}
\end{equation}

Recently, \cite{abbott2021population} estimated the local BNS event rate density as $\dot{\rho}_{0,{\rm BNS}} = 320^{+490}_{-240}\,{\rm Gpc}^{-3}\,{\rm yr}^{-1}$ based on the GW observations during the first half of the third observing (O3) run \citep[see][for a review of $\dot{\rho}_{0,{\rm BNS}}$]{mandel2021}. Hereafter, if not otherwise specified, $\dot{\rho}_{0,{\rm BNS}}$ used in our calculations are simply set as the median value of the GW constraint by the LIGO/Virgo Collaboration (LVC), i.e., $\dot{\rho}_{0,{\rm BNS}} \simeq 320\,{\rm Gpc}^{-3}\,{\rm yr}^{-1}$.

BNS mergers can be thought as occurring with a delay timescale with respect to the star formation history. {{The Gaussian delay model \citep{virgili2011}, log-normal delay model \citep{nakar2006,wanderman2015}, and power-law delay model \citep{virgili2011,hao2013,davanzo2014} are main types of delay time distributions.}} {{\cite{sun2015} suggested that the power-law delay model leads to a wider redshift distribution of BNS merger than other two models, while recent observations on sGRBs by \cite{zevin2022,fong2022,oconnor2022,nugent2022} supported power-law delay more. Although many debates, for simplicity,}} we only adopt the log-normal delay model as our merger delay model and the analytical fitting expression of $f(z)$ is presented as Equation (A8) in \cite{zhu2021kilonova}. With known redshift distribution $f(z)$, we randomly simulate a group of $n_{\rm sim} = 5\times 10^6$ BNS events in the universe based on Equation (\ref{equ:merger_rate}). 
For each BNS event, we then generate its EM emissions. We briefly assume that all of BNS events in the universe would only power three main types of EM signals, i.e. the jet afterglow, the kilonova, and the sGRB. {{We assume that cosmological kilonovae are AT2017gfo-like with the consideration of the viewing-angle effect, while the properties of afterglows are consistent with those of cosmological sGRB afterglows. The modeling details of redshift distribution, jet afterglow and kilonova emissions of BNS mergers has been presented in \citetalias{zhu2021kilonovaafterglow}.}} Our viewing-angle-dependent semianalytical model of sGRB emission follows  \cite{song2019}  and \cite{yu2021}. The signature of sGRBs depends on the on-axis equivalent isotropic energy $E_0$, the core half-opening angle $\theta_{\rm c}$, and the latitudinal viewing angle $\theta_{\rm view}$, while the afterglow emission has a dependence on four additional parameters, i.e. number density of interstellar medium $n$, power-law index of the electron distribution $p$, fractions of shock energy distributed in electrons, $\varepsilon_{e}$, and in magnetic fields, $\varepsilon_{B}$. Furthermore, the kilonova emission is only determined by $\theta_{\rm view}$. According to the distributions of above parameters as described in \citetalias{zhu2021kilonovaafterglow} in detail, one can randomly generate the EM emission components for each simulated BNS event.

\subsection{Classification of Detectable EM Counterparts}

\begin{deluxetable*}{cccc}[tpb!]
\tablecaption{Sample for EM counterparts of BNS mergers} \label{tab:Subsamples}
\tablecolumns{10}
\tablewidth{0pt}
\tablehead{
\colhead{Sample} &
\colhead{sGRB} &
\colhead{kilonova-dominated} &
\colhead{afterglow-dominated} 
}
\startdata
kilonova w/ sGRB & \ding{51} & \ding{51} & \ding{55} \\
kilonova w/o sGRB & \ding{55} & \ding{51} & \ding{55} \\
afterglow w/ sGRB & \ding{51} & \ding{55} & \ding{51} \\
afterglow w/o sGRB & \ding{55} & \ding{55} & \ding{51} \\
sGRB only & \ding{51} & \ding{55} & \ding{55}
\enddata
\end{deluxetable*}

{{We divide the detectable events into two main groups based on the relative brightness of the detected kilonova and afterglow. If the peak kilonova flux is larger than five times of the afterglow flux, i.e., $F_{\nu,{\rm KN}}(t_{\rm KN,p})>5F_{\nu,{\rm AG}}(t_{\rm KN,p})$, where $t_{\rm KN,p}$ is the peak time of the kilonova, we qualify these events into ``kilonova-dominated sample". For such events, kilonova emission at the peak time would be at least two magnitudes brighter than that of the associated afterglow emission, so that this requirement can guarantee a clear kilonova signal for observers. For on-axis or near-on-axis afterglows, some bright kilonovae can appear detectable as an excess flux compared to the afterglow power-law decay, which are also defined as kilonova-dominated events. Other events are classified as ``afterglow-dominated sample", since the observed kilonova signals of these events may be ambiguous. In \citetalias{zhu2021kilonovaafterglow}, we have shown that $\sim50\%$ on-axis and nearly-on-axis afterglows are brighter than the associated kilonovae at the peak time. Thus, most of them would be afterglow-dominated. Only at large viewing angles with $\sin\theta_{\rm v}\gtrsim 0.20$, the EM signals of most BNS mergers would be kilonova-dominated and some off-axis afterglows may emerge at $\sim5-10$\,day after the mergers.}}

{{Some optically-discovered EM counterparts of BNS mergers could be associated with sGRB observations. For the GW-triggered ToO searches, the observations of sGRBs can cooperate on the constraint on the sky location for BNS GW alerts, which would help us find the EM counterparts. {{On the basis of whether or not an sGRB is detected}}, we can further divide each sample into two subsamples, i.e., (1) kilonova-dominated sample: kionova with (w/) sGRB and kilonova without (w/o) sGRB; (2) afterglow-dominated sample: afterglow w/ sGRB and afterglow w/o sGRB. Furthermore, {{a fraction of BNS mergers may only be detected in the $\gamma$-ray band without any detection of an associated optical afterglow or kilonova.}} Thus, we totally define five subsamples for the EM counterparts of BNS mergers (see Table \ref{tab:Subsamples}).}}

{{SGRBs are believed to be triggered if $F_\gamma > F_{\gamma,{\rm limit}}$, where $F_\gamma$ is the $\gamma$-band flux for each BNS GW event \citep[see ][for the details of sGRB model]{song2019,yu2021} and $F_{\gamma,{\rm limit}}$ is the effective sensitivity limit for various $\gamma$-ray detectors. Many GRB detectors with quick response and wide field of view, e.g., Swift \citep{gehrels2004}, AstroSAT \citep{singh2014}, Fermi \citep{meegan2009}, and GECAM \citep{zhang2018,song2019}, will work during O4. In our calculations, we simply set $F_{\gamma,{\rm limit}} \sim 2 \times 10^{-7}\,{\rm erg}\,{\rm s}^{-1}$ in $50-300\,{\rm keV}$ which is the effective sensitivity limit of Fermi-GBM \citep{meegan2009} and GECAM \citep{zhang2018,song2019}, in view of that Fermi-GBM and GECAM can nearly achieve an all-sky coverage to detect GRB events\footnote{{{Compared with Fermi-GBM and GECAM, Swift-BAT \citep{gehrels2004,lien2014} has a much lower sensitivity of $F_{\gamma,{\rm limit}} \sim 1\times10^{-8}\,{\rm erg}\,{\rm s}^{-1}$ in $15-150\,{\rm keV}$. However, unlike Fermi-GBM and GECAM that can nearly achieve an all-sky coverage, the Swift-BAT's FoV is $\sim1.4\,{\rm sr}$. It is expected that the number of events with $\gamma$-ray triggers by Swift/BAT could be even lower than that by Fermi-GMB and GECAM due to its limited FoV \citep[e.g.,][]{song2019}.}} }}}.

\section{Detectability for Serendipitous Searches \label{sec:serendiption}}

{{In this Section, we will introduce the method for the calculations of the EM detection rate via the serendipitous observations, investigate on the optimal search strategy, and show our simulated optically-discovered detection rates of the kilonova-dominated and afterglow-dominated events for some specific survey projects. By considering the observations of sGRBs, the population properties for detectable EM events via the serendipitous searches are detailedly discussed in the following.}}

\subsection{Method} \label{sec:serendiption_method}

Following \cite{zhu2021kilonova}, we adopt a method of probabilistic statistical analysis to estimate the EM detection rate for BNS mergers. The probability that a single simulated event can be detected could be considered as the ratio of survey area within the time duration ($\Delta t$) that the brightness of the associated EM signal is above the limiting magnitude ($m_{\rm limit}$) to the area of the celestial sphere ($\Omega_{\rm sph} = 41252.96\,{\rm deg}^2$). The maximum probability for a source to be detected is $\Omega_{\rm FoV}\dot{t}_{\rm ope}\Delta t/\Omega_{\rm sph}(n_{\rm exp} t_{\rm exp} + t_{\rm oth})$, where $\Omega_{\rm FoV}$ is the field of view (FoV) for the specific survey project, $\dot{t}_{\rm ope}$ is the average operation time per day, $n_{\rm exp}$ is defined as the exposure number for each visit, $t_{\rm exp}$ is the exposure time, and $t_{\rm other}$ is other time spent for each visit. However, high-cadence observations would restrict the survey area, which means that the probability of a source being detected by the high-cadence search would be a constant, i.e., $\Omega_{\rm FoV}\dot{t}_{\rm ope}t_{\rm cad}/\Omega_{\rm sph}(n_{\rm exp}t_{\rm exp} + t_{\rm oth})$, where the cadence time $t_{\rm cad}$ defined as the interval between consecutive observations of the same sky area by a telescope. Furthermore, the event should appear in the sky coverage of the survey telescope that one can have a chance to discover it. Thus, we simply set an upper limit on the probability for a source that can be detected, which is expressed as $\Omega_{\rm cov}/\Omega_{\rm sph}$ with $\Omega_{\rm cov}$ being the detectable sky coverage for a specific survey project. By counting the detection probabilities of all simulated events, one can write the EM detection rate for the serendipitous observations as 
 
\begin{equation}
\label{equ:EMdetection_rate}
    \dot{N}_{\rm EM} \approx \frac{\dot{N}_{\rm BNS}}{n_{\rm sim}}\sum_{i = 1}^{n_{\rm sim}} \min\left[\frac{\Omega_{\rm cov}}{\Omega_{\rm sph}} , \frac{\Omega_{\rm FoV}\dot{t}_{\rm ope}\min(t_{\rm cad},\Delta t_i)}{\Omega_{\rm sph}(n_{\rm exp}t_{\rm exp} + t_{\rm oth})}\right],
\end{equation}
We roughly assume that the average operation time per day is $\dot{t}_{\rm ope} \approx 6\,{\rm hr}\,{\rm day}^{-1}$ for all survey projects except for SiTian. The time spent for each visit ${t}_{\rm oth}$ is dependent on the technical performance of specific survey project and different search strategy. Because $t_{\rm oth}$ is uncertain, we set it as a constant for each survey project, i.e., $t_{\rm oth} = 15\,{\rm s}$. 

\begin{deluxetable*}{cccccccccccc}[htpb]
\tablecaption{Summary Technical Information for Each Survey} \label{tab:SurveyProject}
\tablewidth{0pt}
\tablehead{
\colhead{Telescope} &
\multicolumn{6}{c}{$m_{\rm limit} = a\times t_{\rm exp}^b$} &
\colhead{{$t_{\rm exp}/{\rm s}$}} & 
\colhead{{$m_{g,{\rm limit}}/{\rm mag}$}} & 
\colhead{FoV/${\rm deg}^2$} &
\colhead{Sky Coverage/${\rm deg}^2$} &
\colhead{Reference}
}
\startdata
\multirow{3}{*}{ZTF}& \multicolumn{2}{c}{$g$}   & \multicolumn{2}{c}{$r$}   & \multicolumn{2}{c}{$i$} & 30 & 20.3 & \multirow{3}{*}{47.7} & \multirow{3}{*}{30,000}     & \multirow{3}{*}{(1)} \\  & \multicolumn{2}{c}{18.62} & \multicolumn{2}{c}{18.37} & \multicolumn{2}{c}{17.91} & 180 & 21.3 &     & \\
  & \multicolumn{2}{c}{0.026} & \multicolumn{2}{c}{0.026} & \multicolumn{2}{c}{0.027} &   300 & 21.6  &     & \\ \hline
\multirow{3}{*}{Mephisto} & $u$   & $v$   & $g$   & $r$   & $i$    & $z$  & 30 & 22.4 & \multirow{3}{*}{3.14} & \multirow{3}{*}{26,000}     & \multirow{3}{*}{(2)} \\
  & 18.45 & 18.54 & 19.91 & 19.91 & 19.68  & 18.71&  180 & 23.8   &     & \\
  & 0.043 & 0.042 & 0.034 & 0.032 & 0.030  & 0.033&  300 & 24.2   &     & \\ \hline
\multirow{3}{*}{WFST}     & $u$   & $g$   & $r$   & $i$   & $z$    & $w$  & 30 & 23.0 & \multirow{3}{*}{6.55} & \multirow{3}{*}{20,000}     & \multirow{3}{*}{(3)} \\
  & 20.70 & 21.33 & 21.13 & 20.46 & 19.41  & 21.33&  180 &  23.9 &   &     & \\
  & 0.022 & 0.022 & 0.022 & 0.023 & 0.024  & 0.022&  300 & 24.2 &  &     & \\ \hline
\multirow{3}{*}{LSST}     & $u$   & $g$   & $r$   & $i$   & $z$    & $y$  & 30 & 25.1 & \multirow{3}{*}{9.6}  & \multirow{3}{*}{20,000}     & \multirow{3}{*}{(4)} \\
  & 22.03 & 23.60 & 22.54 & 21.73 & 21.83 & 21.68&  180 & 25.9 &   &     & \\
  & 0.025 & 0.018 & 0.023 & 0.030 & 0.031  & 0.032&  300 & 26.2 &  &     & \\ \hline
\multirow{3}{*}{SiTian\tablenotemark{{\rm *}}}   & \multicolumn{2}{c}{$g$}   & \multicolumn{2}{c}{$r$}   & \multicolumn{2}{c}{$i$}  & 30 & 20.3  & \multirow{3}{*}{600}  & \multirow{3}{*}{30,000}     & \multirow{3}{*}{(5)} \multirow{3}{*}{} \\
  & \multicolumn{2}{c}{18.62} & \multicolumn{2}{c}{18.37} & \multicolumn{2}{c}{17.91} &  180 & 21.3 &   &     & \\
  & \multicolumn{2}{c}{0.026} & \multicolumn{2}{c}{0.026} & \multicolumn{2}{c}{0.027} &   300 & 21.6 &  &     &   
\enddata
\tablecomments{The columns are [1] the survey project; [2] the search limiting magnitude $m_{\rm limit}$ as a logarithmic function of exposure time $t_{\rm exp}$ in different bands for specific survey project (parameter $a$ and $b$ are respectively the values at the second and third sub-rows of each row); 
{{[3] exposure time $t_{\rm exp}$; [4] $g$-band limiting magnitude $m_{g,\rm{limit}}$ corresponding to different exposure times;}} [5] field of view $\Omega_{\rm FoV}$; [6] detectable sky coverage $\Omega_{\rm cov}$; [7] references. \\Reference: (1) \cite{bellm2019,masci2019}; (2) Er et al. (2022), in preparation; \cite{lei2021} (3) Kong et al. (2022), in preparation; \cite{shi2018} (4) \cite{lsst2009}; (5) \cite{liu2021}.}
\tablenotetext{*}{ The technical specification of the limiting magnitude in the $g$-band stacked images for SiTian is similar to that for ZTF \citep{liu2021}. SiTian would simultaneously observe the same visit in three different filters ($u$, $g$, $i$). Due to the lack of the technical information in $u$ and $i$ band of SiTian, we simply use the technical information of ZTF in $gri$ bands to calculate the EM detection rates by SiTian. }
\end{deluxetable*}

In order to reject the supernova background and other rapid-evolving transients, in \citetalias{zhu2021kilonovaafterglow}, we showed that one can use the unique color evolution of kilonovae and afterglows to identify them among the observed transients. We require that the judgement condition for the detection of the kilonova and/or afterglow by a serendipitous search is that ``two different exposure filters have at least two detection epochs". It would be $n_{\rm exp} = 1$ for Mephisto and SiTian since these two survey projects can achieve simultaneous imaging in three bands,\footnote{{{The optical system of Mephisto consists of a modified Ritchey-Chr\'{e}tien design with three refractive correctors and three cubes for beam splitting so that it can be capable of simultaneously imaging the same patch of sky in three bands (Er et al. 2022, in preparation). SiTian is composed of a number of ``units" which is made of three 1-m-class Schmidt telescopes (see Section \ref{sec:sitian} for more details). Both of them can achieve simultaneous imaging in three bands.}}} while $n_{\rm exp} = 2$ for ZTF, WFST, and LSST.

{{The values of some technical parameters, including the expected limiting magnitude which is a logarithmic function of exposure time in each band, FoV, detectable sky coverage, for the survey telescopes we considered are presented in Table \ref{tab:SurveyProject}. As examples, we also list $g$-band limiting magnitudes $m_{g,{\rm limit}}$ with different exposure times of $t_{\rm{exp}} = 30,\,180,\,300\,{\rm s}$ for these survey telescopes in Table \ref{tab:SurveyProject}. Thus, survey telescopes with aperture that smaller than ZTF and SiTian can have a limiting magnitude of $m_{\rm limit}\lesssim20\,{\rm mag}$. The detection depth of ZTF and SiTian locates in a range of $20\,{\rm mag}\lesssim{m}_{\rm limit}\lesssim22\,{\rm mag}$. It expects that $22\,{\rm mag}\lesssim{m}_{\rm {limit}}\lesssim24\,{\rm mag}$ applies to Mephisto and WFST. $24\lesssim{m}_{\rm {limit}}\lesssim26\,{\rm mag}$ can be only achieved by LSST.   }}

\subsection{$\Delta t_{\rm EM}$ and Cadence Time Selection} 

\begin{deluxetable*}{ccccccccccc}[htpb]
\tablecaption{Time during which the brightness of EM counterpart is above the limiting magnitude} \label{tab:Delta_t}
\tablehead{
\colhead{Filter} &
\colhead{Parameter} &
\colhead{$m_{\rm limit} = 18\,{\rm mag}$} &
\colhead{$19\,{\rm mag}$} &
\colhead{$20\,{\rm mag}$} &
\colhead{$21\,{\rm mag}$} &
\colhead{$22\,{\rm mag}$} &
\colhead{$23\,{\rm mag}$} &
\colhead{$24\,{\rm mag}$} &
\colhead{$25\,{\rm mag}$} &
\colhead{$26\,{\rm mag}$} 
}
\startdata
\multirow{2}{*}{$u$} & $\Delta t_{\rm KN}$ & -- & $0.67^{+0}_{-0.67}$ & $0.65^{+0.77}_{-0.65}$ & $0.82^{+0.63}_{-0.72}$ & $0.8^{+1.1}_{-0.6}$ & $0.7^{+1.4}_{-0.6}$ & $0.7^{+1.3}_{-0.5}$ & $0.7^{+1.2}_{-0.6}$ & $0.8^{+1.2}_{-0.6}$ \\
 & $\Delta t_{\rm AG}$ & $0.12^{+0.55}_{-0.11}$ & $0.14^{+0.68}_{-0.12}$ & $0.18^{+0.91}_{-0.16}$ & $0.2^{+1.3}_{-0.2}$ & $0.3^{+1.9}_{-0.3}$ & $0.4^{+3.0}_{-0.4}$ & $0.6^{+4.7}_{-0.5}$ & $0.9^{+7.5}_{-0.8}$ & $2^{+12}_{-1}$ \\ \hline
\multirow{2}{*}{$g$} & $\Delta t_{\rm KN}$ & $0.83^{+0}_{-0.83}$ & $0.7^{+1.1}_{-0.7}$ & $1.3^{+0.8}_{-1.2}$ & $1.2^{+1.4}_{-1.0}$ & $1.2^{+1.9}_{-0.9}$ & $1.1^{+1.7}_{-0.8}$ & $1.0^{+1.7}_{-0.7}$ & $0.9^{+1.8}_{-0.7}$ & $0.9^{+1.9}_{-0.7}$ \\
 & $\Delta t_{\rm AG}$ & $0.12^{+0.58}_{-0.11}$ & $0.15^{+0.73}_{-0.13}$ & $0.19^{+0.97}_{-0.17}$ & $0.2^{+1.4}_{-0.2}$ & $0.3^{+2.1}_{-0.3}$ & $0.4^{+3.3}_{-0.4}$ & $0.6^{+5.2}_{-0.5}$ & $1.0^{+8.1}_{-0.9}$ & $2^{+13}_{-2}$ \\ \hline
\multirow{2}{*}{$v$} & $\Delta t_{\rm KN}$ & $1.4^{+0}_{-1.4}$ & $0.7^{+1.1}_{-0.6}$ & $1.1^{+1.5}_{-0.8}$ & $1.0^{+2.4}_{-0.8}$ & $1.1^{+2.3}_{-0.7}$ & $1.0^{+2.0}_{-0.8}$ & $1.1^{+1.9}_{-0.8}$ & $1.2^{+2.0}_{-0.9}$ & $1.3^{+1.9}_{-0.7}$ \\
 & $\Delta t_{\rm AG}$ & $0.13^{+0.60}_{-0.12}$ & $0.15^{+0.77}_{-0.13}$ & $0.2^{+1.0}_{-0.2}$ & $0.2^{+1.5}_{-0.2}$ & $0.3^{+2.2}_{-0.3}$ & $0.4^{+3.4}_{-0.4}$ & $0.6^{+5.4}_{-0.6}$ & $1.0^{+8.6}_{-1.0}$ & $2^{+13}_{-2}$ \\ \hline
\multirow{2}{*}{$w$} & $\Delta t_{\rm KN}$ & $0.5^{+1.1}_{-0.5}$ & $0.89^{+0.88}_{-0.60}$ & $1.2^{+1.6}_{-0.9}$ & $1.2^{+2.3}_{-0.8}$ & $1.2^{+2.3}_{-0.8}$ & $1.1^{+2.3}_{-0.8}$ & $1.1^{+2.2}_{-0.8}$ & $1.2^{+2.1}_{-0.8}$ & $1.3^{+2.0}_{-1.0}$ \\
 & $\Delta t_{\rm AG}$ & $0.13^{+0.61}_{-0.12}$ & $0.15^{+0.78}_{-0.13}$ & $0.2^{+1.0}_{-0.2}$ & $0.2^{+1.5}_{-0.2}$ & $0.3^{+2.2}_{-0.3}$ & $0.4^{+3.5}_{-0.4}$ & $0.6^{+5.5}_{-0.6}$ & $1.1^{+8.7}_{-1.0}$ & $2^{+14}_{-2}$ \\ \hline
\multirow{2}{*}{$r$} & $\Delta t_{\rm KN}$ & $0.7^{+1.0}_{-0.7}$ & $1.07^{+0.88}_{-0.69}$ & $1.3^{+1.7}_{-1.0}$ & $1.2^{+2.5}_{-0.8}$ & $1.1^{+2.4}_{-0.8}$ & $1.3^{+2.1}_{-0.9}$ & $1.3^{+2.2}_{-0.9}$ & $1.2^{+2.4}_{-0.9}$ & $1.3^{+2.4}_{-0.9}$ \\
 & $\Delta t_{\rm AG}$ & $0.13^{+0.63}_{-0.12}$ & $0.15^{+0.79}_{-0.13}$ & $0.2^{+1.1}_{-0.2}$ & $0.3^{+1.5}_{-0.2}$ & $0.3^{+2.3}_{-0.3}$ & $0.4^{+3.6}_{-0.4}$ & $0.7^{+5.6}_{-0.6}$ & $1.1^{+8.9}_{-1.0}$ & $2^{+14}_{-2}$ \\ \hline
\multirow{2}{*}{$i$} & $\Delta t_{\rm KN}$ & $0.7^{+1.9}_{-0.7}$ & $1.3^{+1.3}_{-0.7}$ & $1.5^{+2.2}_{-1.3}$ & $1.4^{+3.1}_{-0.9}$ & $1.3^{+2.6}_{-1.0}$ & $1.3^{+2.4}_{-1.0}$ & $1.4^{+2.4}_{-0.9}$ & $1.4^{+2.5}_{-1.0}$ & $1.7^{+2.6}_{-1.2}$ \\
 & $\Delta t_{\rm AG}$ & $0.13^{+0.67}_{-0.12}$ & $0.16^{+0.84}_{-0.14}$ & $0.2^{+1.1}_{-0.2}$ & $0.3^{+1.6}_{-0.2}$ & $0.3^{+2.4}_{-0.3}$ & $0.5^{+3.8}_{-0.4}$ & $0.7^{+6.0}_{-0.6}$ & $1.1^{+9.4}_{-1.1}$ & $2^{+15}_{-2}$ \\ \hline
\multirow{2}{*}{$z$} & $\Delta t_{\rm KN}$ & $0.8^{+2.3}_{-0.8}$ & $1.6^{+1.6}_{-1.1}$ & $2.0^{+2.3}_{-1.4}$ & $1.8^{+3.0}_{-1.3}$ & $1.8^{+2.7}_{-1.3}$ & $1.8^{+2.6}_{-1.3}$ & $1.6^{+2.9}_{-1.1}$ & $1.6^{+3.1}_{-1.2}$ & $1.7^{+2.8}_{-1.3}$ \\
 & $\Delta t_{\rm AG}$ & $0.14^{+0.69}_{-0.13}$ & $0.16^{+0.89}_{-0.14}$ & $0.2^{+1.2}_{-0.2}$ & $0.3^{+1.7}_{-0.2}$ & $0.4^{+2.6}_{-0.3}$ & $0.5^{+4.0}_{-0.4}$ & $0.7^{+6.3}_{-0.7}$ & $1^{+10}_{-1}$ & $2^{+15}_{-2}$ \\ \hline
\multirow{2}{*}{$y$} & $\Delta t_{\rm KN}$ & $1.0^{+2.2}_{-1.0}$ & $1.8^{+1.5}_{-1.7}$ & $2.1^{+2.2}_{-1.7}$ & $1.8^{+3.6}_{-1.4}$ & $1.9^{+2.9}_{-1.4}$ & $1.9^{+2.7}_{-1.4}$ & $1.9^{+2.9}_{-1.4}$ & $1.9^{+2.8}_{-1.5}$ & $1.7^{+3.2}_{-1.4}$ \\
 & $\Delta t_{\rm AG}$ & $0.14^{+0.71}_{-0.12}$ & $0.17^{+0.90}_{-0.15}$ & $0.2^{+1.2}_{-0.2}$ & $0.3^{+1.8}_{-0.2}$ & $0.4^{+2.6}_{-0.3}$ & $0.5^{+4.1}_{-0.5}$ & $0.8^{+6.5}_{-0.7}$ & $1^{+10}_{-1}$ & $2^{+16}_{-2}$
\enddata
\tablecomments{The values are the timescales during which the brightness of associated kilonova ($\Delta t_{\rm KN}$) and afterglow ($\Delta t_{\rm AG}$) is above the $5\sigma$ limiting magnitude in different bands with $90\%$ interval. }
\end{deluxetable*}

\begin{figure}
   \centering
    \includegraphics[width = 0.99\linewidth , trim = 80 120 95 35, clip]{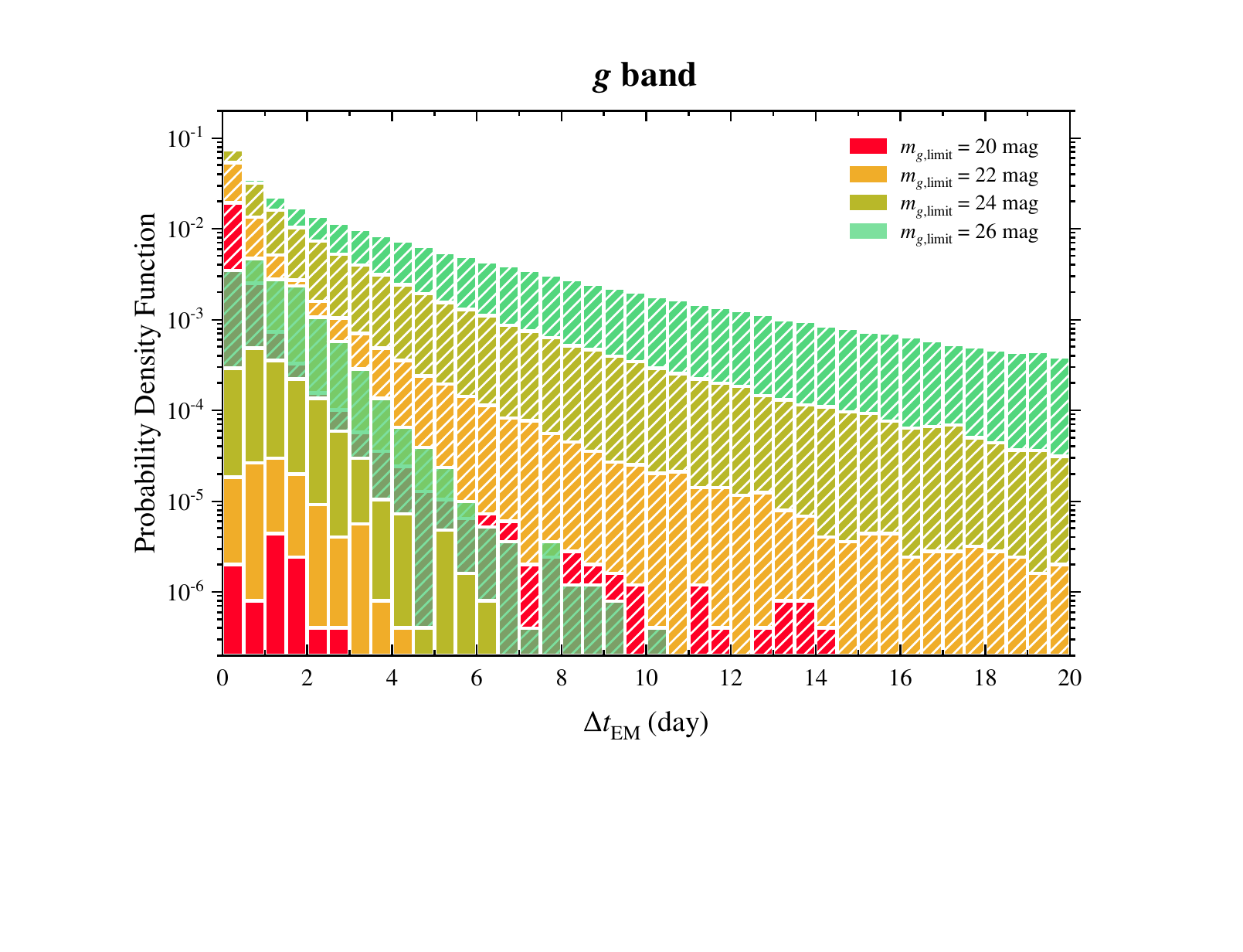}
    \includegraphics[width = 0.99\linewidth , trim = 80 120 95 35, clip]{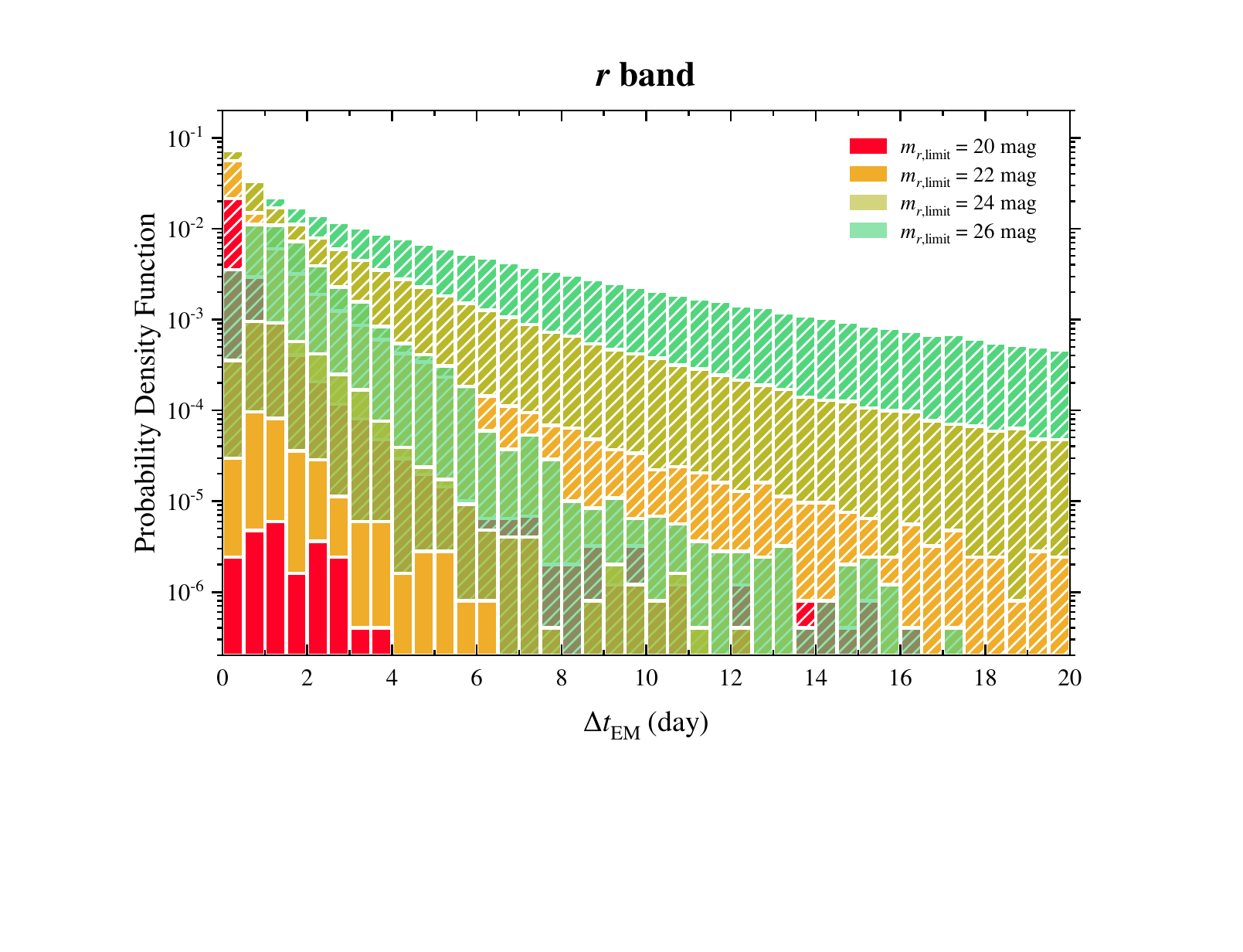}
    \includegraphics[width = 0.99\linewidth , trim = 80 120 95 35, clip]{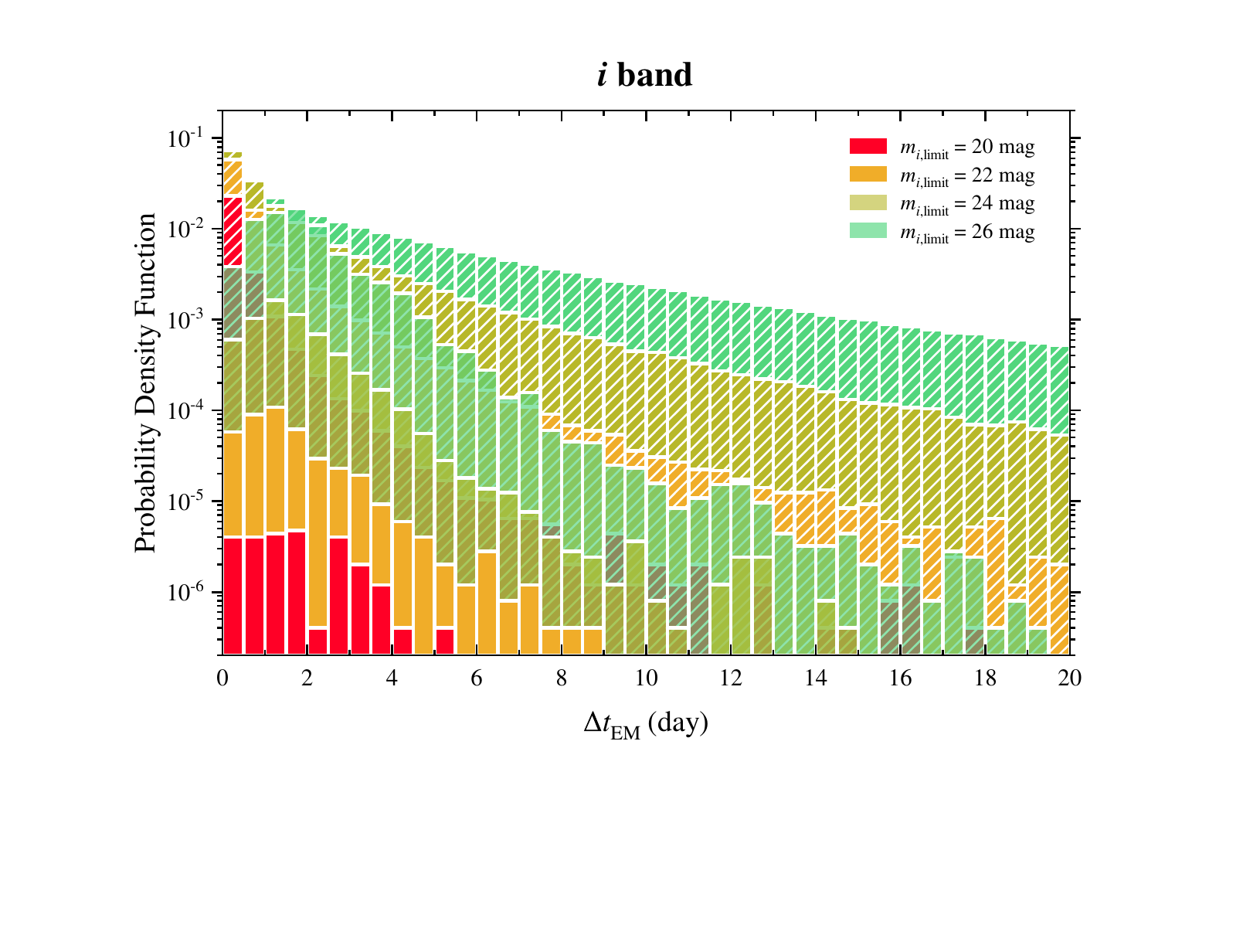}
    \caption{Crimson, orange, yellowgreen, and green histograms are the probability density functions of $\Delta t_{\rm KN}$ (solid histograms) and $\Delta t_{\rm AG}$ (striped histograms) for a limiting magnitude of $m_{\rm limit} = 20,\,22,\,24,\,{\rm and}\ 26\,{\rm mag}$, in $g$ band (top panel), $r$ band (medium panel), and $i$ band (bottom panel). The bin width of the histograms is set as $\Delta = 0.5\,{\rm day}$.}
    \label{fig:Delta_t}
\end{figure}

As listed in Table \ref{tab:Delta_t}, we show the $90\%$ credible regions of two timescales, i.e., $\Delta t_{\rm KN}$ and $\Delta t_{\rm AG}$, with different filters and different limiting magnitudes. These two parameters are respectively defined as the timescales during which the brightness of the associated kilonova and afterglow is above the limiting magnitude in different bands. Because $gri$ bands are the common filters used by various survey projects, we only show the probability density functions of $\Delta t_{\rm KN}$ and $\Delta t_{\rm AG}$ with different searching magnitudes in these three bands in Figure \ref{fig:Delta_t}.

{{As shown in Table \ref{tab:Delta_t}}}, for a limiting magnitude of $m_{\rm limit}\leq 19\,{\rm mag}$, the values of $\Delta t_{\rm KN}$ may be imprecise, due to the limited amount of the available data. {{For $m_{\rm limit}\geq 20\,{\rm mag}$, }} one can see that the median value of $\Delta t_{\rm KN}$ is referred to lie $\sim 0.6-1.4\,{\rm day}$ in optical and $\sim1.4-2.1\,{\rm day}$ in infrared, which may be uncorrelated with the limiting magnitude $m_{\rm limit}$. If the observer wants to achieve at least two detection epochs for at least $50\%$ of the observable kilonova signals, the cadence time $t_{\rm cad}$ should be less than half of the median value of $\Delta t_{\rm KN}$. This means $t_{\rm cad}$ should be $t_{\rm cad}\lesssim 0.3-0.7\,{\rm day}$ if one uses an optical band to search for kilonovae and $t_{\rm cad}\lesssim 0.7-1.0\,{\rm day}$ by an infrared band {{for all survey projects}}. 

Unlike $\Delta t_{\rm KN}$, there exists a positive correlation between $\Delta t_{\rm AG}$ and $m_{\rm limit}$ {{listed in Table \ref{tab:Delta_t}.}}. The median value of $\Delta t_{\rm KN}$ would be always larger than $\Delta t_{\rm AG}$ if $m_{\rm limit}\lesssim 24\,{\rm mag}$. {{Thus, LSST, which has a limiting magnitude of $m_{\rm {limit}}\gtrsim24\,{\rm mag}$, can find $\gtrsim50\%$ detectable afterglows brighter than the searching limiting magnitude by adopting a cadence to search for kilonovae.}} As shown in Figure \ref{fig:Delta_t}, the probability density function of $\Delta t_{\rm AG}$ is significantly higher than that of $\Delta t_{\rm KN}$, especially for searching with a relatively shallow limiting magnitude in a bluer filter band. Thus, it may be easier to discover optical afterglows by adopting the cadence of searching for kilonovae. 

\subsection{Optimal Search Strategy}

\begin{figure*}[htpb]
    \centering
    \includegraphics[width = 0.99\linewidth , trim = 0 50 0 0, clip]{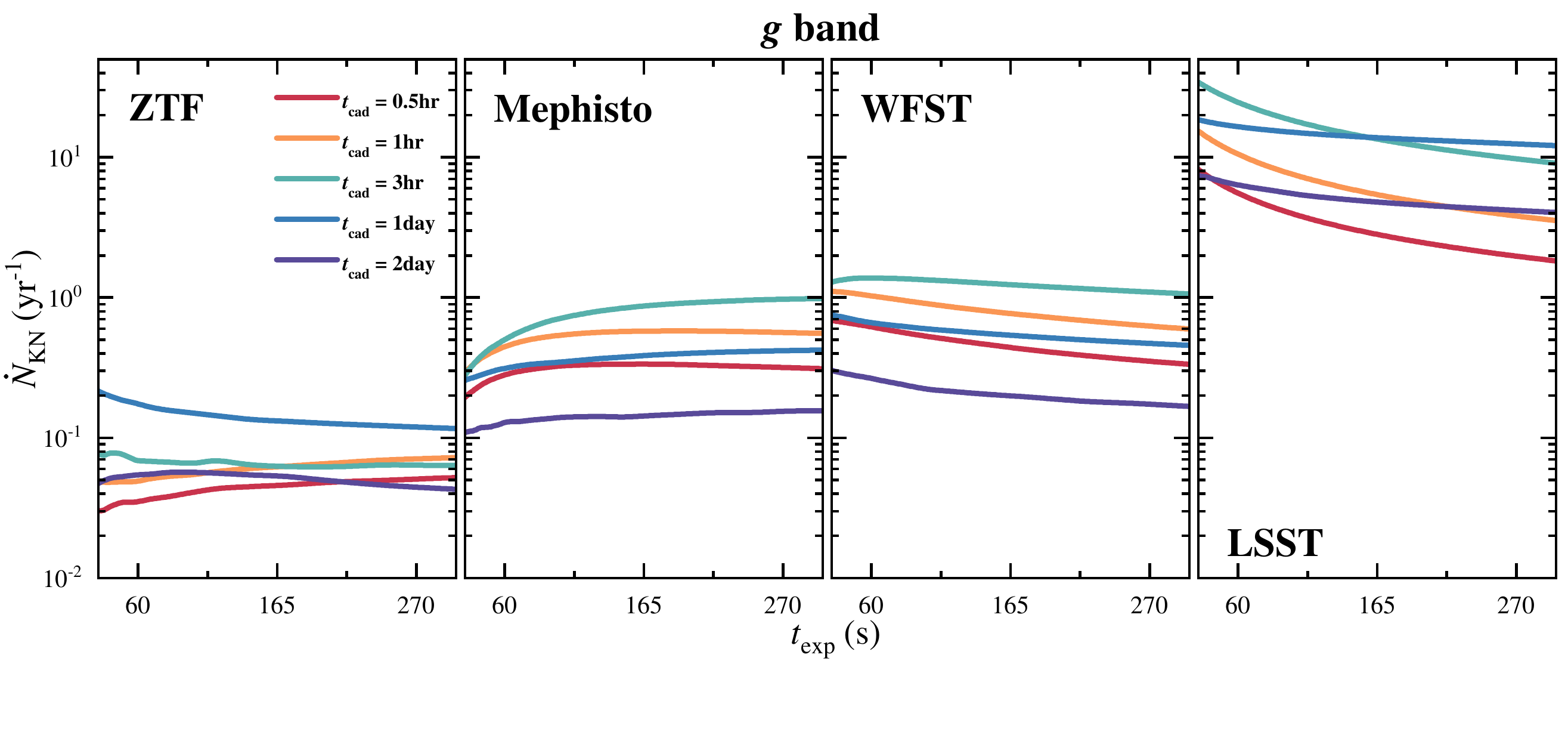}
    \includegraphics[width = 0.99\linewidth , trim = 0 50 0 0, clip]{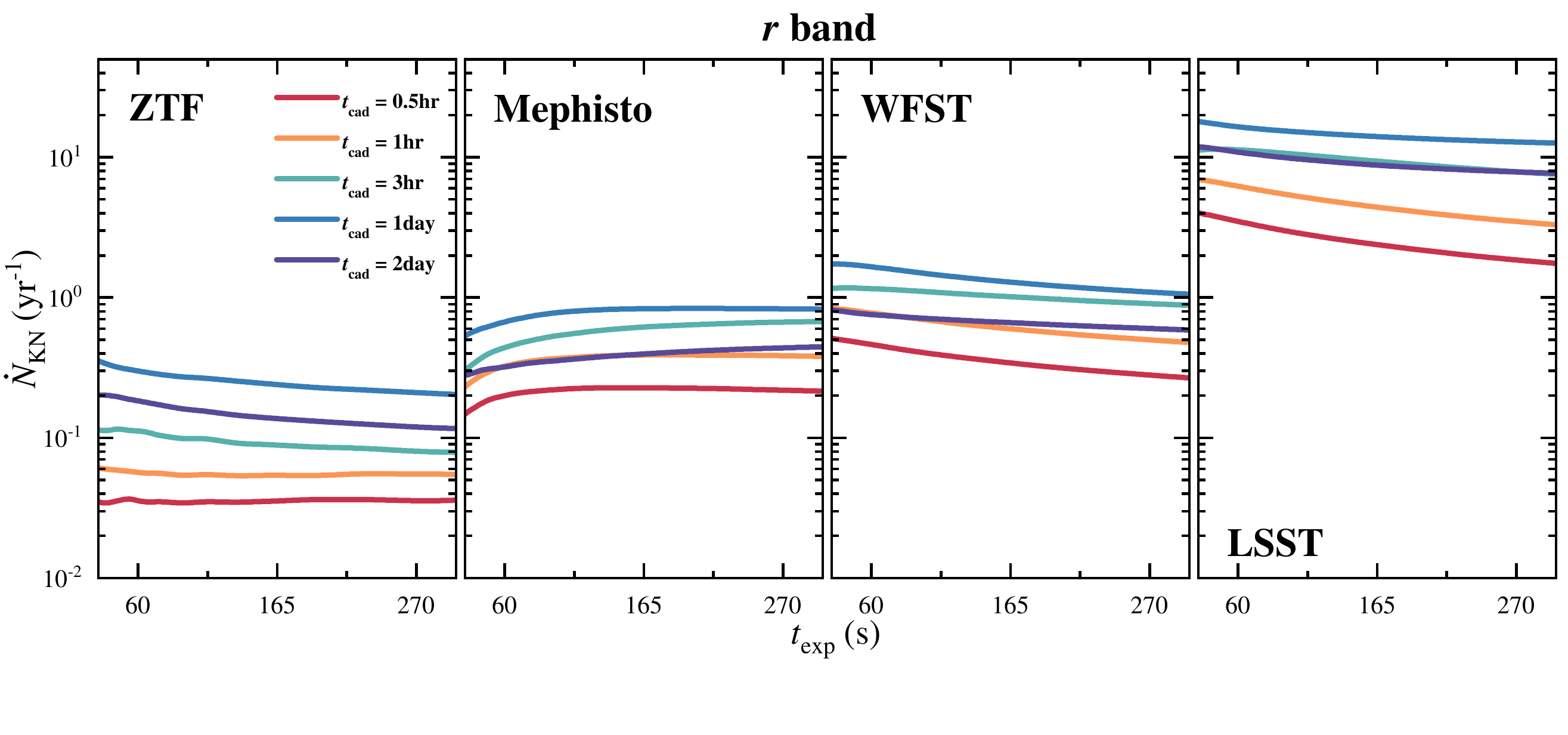}
    \includegraphics[width = 0.99\linewidth , trim = 0 50 0 0, clip]{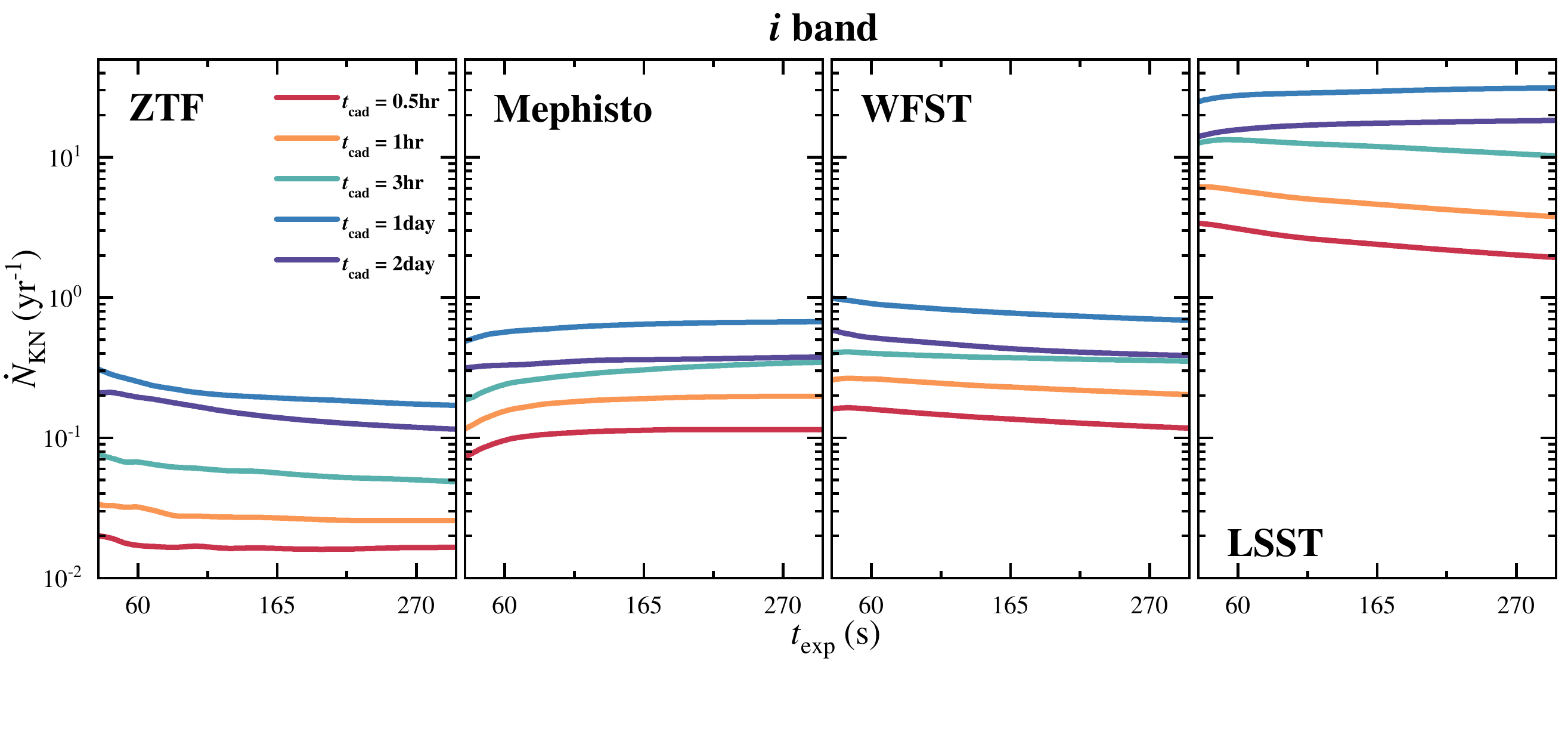}
    \caption{Detection rates of kilonova-dominated sample as functions of exposure time $t_{\rm exp}$ and cadence time $t_{\rm cad}$ for serendipitous observations. Four survey projects, including ZTF, Mephisto, WFST, and LSST (from left to right panels), are considered. The panels from top to bottom represent events of kilonova-dominated sample to be detected in the $g$ band, $r$ band, and $i$ band, respectively. Red, orange, green, blue, and violet lines are the detection rates by adopting cadence searching strategies of $t_{\rm cad} = 0.5\,{\rm hr},\,1\,{\rm hr},\,3\,{\rm hr},\,1\,{\rm day},\,{\rm and}\ 2\,{\rm day}$, respectively.   }
    \label{fig:Blind_KN}
\end{figure*}

\begin{figure*}[htpb]
    \centering
    \includegraphics[width = 0.99\linewidth , trim = 0 50 0 35, clip]{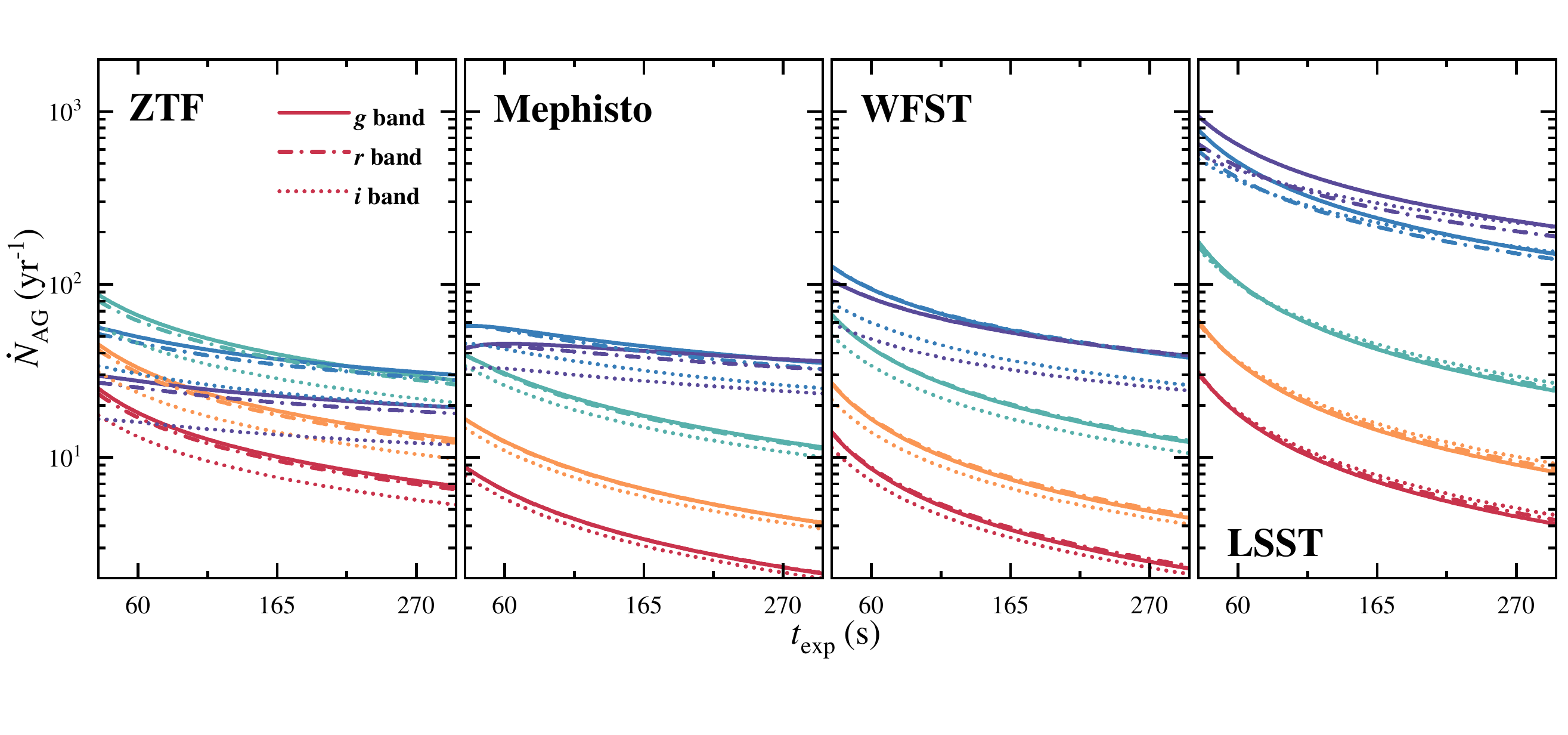}
    \caption{Similar to Figure \ref{fig:Blind_KN}, but for the detection rates of afterglow-dominated sample. Solid, dashed-doted, and dotted lines represent the detection rates in $g$, $r$, and $i$ band, respectively.}
    \label{fig:Blind_AG}
\end{figure*}

We respectively show the detection rates of kilonova-dominated and afterglow-dominated sample for ZTF, Mephisto, WFST, and LSST in Figure \ref{fig:Blind_KN} and Figure \ref{fig:Blind_AG}, by considering exposure time from $30\,{\rm s}$ to $300\,{\rm s}$ and five different cadence timescales $t_{\rm cad} = 0.5\,{\rm hr},\,1\,{\rm hr},\,3\,{\rm hr},\,1\,{\rm day},\,{\rm and}\ 2\,{\rm day}$. The results shown in Figure \ref{fig:Blind_KN} and Figure \ref{fig:Blind_AG} are only considered in $gri$ bands since these three bands are commonly used for these survey telescopes. Because SiTian is an integrated network of dozens of survey and follow-up telescopes, its survey strategy should have a large difference with that of other survey projects. We will give an separate calculation of EM detection rates for SiTian in Section \ref{sec:sitian}.

As shown in Figure \ref{fig:Blind_KN}, for each survey project, the difference of the detection rate for kilonova-dominated events between different bands seems very small, which is a factor of the order of unity. {{For the cadence choice, we find that}} an one-day cadence strategy can always discover the highest number of kilonovae. {{For the choice of exposure time,}} kilonova detection rates {{by ZTF, WFST, and LSST}} would decline as the exposure time increases. On the contrary, a longer exposure time can discover more kilonova events for Mephisto {{, although the increase in the amount of discovered kilonovae with longer exposure times is not significant.}} Simultaneous imaging in three bands by Mephisto is the reason for the difference of the detection rates between Mephisto and other survey projects. To sum up, an one-day cadence strategy with a $\sim 30\,{\rm s}$ exposure time is recommended to achieve  optimal search for kilonovae. Based on Figure \ref{fig:Blind_KN}, the maximum kilonova detection rates for ZTF, Mephisto, WFST, and LSST are $\sim 0.3\,{\rm yr}^{-1}$, $\sim 0.6\,{\rm yr}^{-1}$, $\sim 1\,{\rm yr}^{-1}$, and $20\,{\rm yr}^{-1}$, respectively.

For afterglow-dominated events, there is no significant difference between searching in the optical and in the infrared bands for each survey project. The detection rates would drop with the increase of the exposure time. By adopting the optimal search strategy for kilonovae, one can also discover many afterglows from BNS mergers whose detection rate is much higher than that of kilonovae. For this case, the afterglow detection rates for ZTF, Mephisto, WFST, and LSST are $\sim 50\,{\rm yr}^{-1}$, $\sim 60\,{\rm yr}^{-1}$, $\sim 100\,{\rm yr}^{-1}$, and $\sim800\,{\rm yr}^{-1}$, respectively.

\subsection{Optimal Search Strategy for SiTian \label{sec:sitian}}

\begin{table*}[htpb] 
\centering
\caption{EM Detection Rates for SiTian \label{tab:SiTian}}
\begin{tabular}{ccccccccccc}
\hline\hline
\multirow{2}{*}{$t_{\rm exp}/{\rm s}$} & \multirow{2}{*}{$\Omega_{\rm FoV,1}/{\rm deg}^2$} & \multirow{2}{*}{$t_{\rm cad,1}/{\rm min}$} & \multirow{2}{*}{$\Omega_{\rm FoV,2}/{\rm deg}^2$} & \multirow{2}{*}{$t_{\rm cad,2}/{\rm min}$} &  & {$\dot{N}_{\rm KN}/{\rm yr}^{-1}$} &  & & {$\dot{N}_{\rm AG}\times10^3/{\rm yr}^{-1}$} &  \\ \cline{6-11} 
 &  &  &  &  & $g$ & $r$ & $i$ & $g$ & $r$ & $i$ \\ \hline
45 (fiducial) & \multirow{4}{*}{350} & 30 & \multirow{4}{*}{250} & 80 & 2.0 & 3.4 & 3.6 & 1.5 & 1.7 & 1.9 \\
75  &  & 45 &  & 120 & 2.4 & 4.3 & 4.7 & 1.7 & 1.9 & 2.0 \\
105 &  & 60 &  & 160 & 2.7 & 5.0 & 5.6 & 1.7 & 1.9 & 2.1 \\
165 &  & 90 &  & 240 & 3.0 & 5.8 & 7.1 & 1.7 & 1.9 & 2.0 \\
\hline
\end{tabular}
\tablecomments{We assume that the time between two visits is $t_{\rm oth} = 15\,{\rm s}$. The operation times for units in China and outside China are assumed to be $\dot{t}_{\rm ope} = 8\,{\rm hr}\,{\rm day}^{-1}$ and $16\,{\rm hr}\,{\rm day}^{-1}$, respectively. The columns are [1] the exposure time; [2] the total field of view of SiTian units in China; [3] the corresponding cadence time for SiTian units in China; [4]  the total field of view of SiTian units outside China; [5] the corresponding cadence time for SiTian units outside China; [6-8] the detection rate of kilonova-dominated events in $gri$ bands; [9-11] the detection rate of afterglow-dominated events in $gri$ bands. }
\end{table*}

SiTian \citep{liu2021,yang2022} is composed of a number of “units” deployed partly in China and partly at various sites around the world. Each unit includes three 1-m-class Schmidt telescopes with a FoV of $\Omega_{\rm FoV} = 25\,{\rm deg}^2$, which will simultaneously observe the same visit in three different optical filters. There will be also three or four 4-m-class telescopes for spectral identification and follow-up studies within the project. 

SiTian {{at full design}} will scan at least $10,000\,{\rm deg}^2$ of sky every $30\,{\rm min}$, down to a detection limit of $g\approx21\,{\rm mag}$ with an exposure time of $\sim 1\,{\rm min}$ using at least $14$ units in China. Furthermore, at least $10$ units outside China can survey an additional $\sim20,000\,{\rm deg}^2$ with a slightly lower cadence (a few hr). Based on this fiducial search plan of SiTian, we change the cadence time and exposure time for all of units to explore the optical search strategy of SiTian, while preserving the same sky coverage. 

The results of the EM detection rates for SiTian are shown in Table \ref{tab:SiTian}. We show that SiTian can detect $\sim2\times10^3\,{\rm yr}^{-1}$ afterglow-dominated events. The detection rate of kilonova-dominated events is $\sim (2-4)\,{\rm yr}^{-1}$ by adopting the fiducial search plan of SiTian. Since kilonovae are very faint, a better search strategy would be to increase the exposure time of the telescopes with the expense of losing the cadence. The detection rate of kilonova-dominated events would slightly rise to $\sim (3-7)\,{\rm yr}^{-1}$ if an exposure time of $165\,{\rm s}$ is used. 

\subsection{{{Population Properties of Detectable EM Events via the Serendipitous Observations}}}

\begin{figure*}[htpb]
    \centering
    \includegraphics[width = 0.99\linewidth , trim = 8 10 10 40, clip]{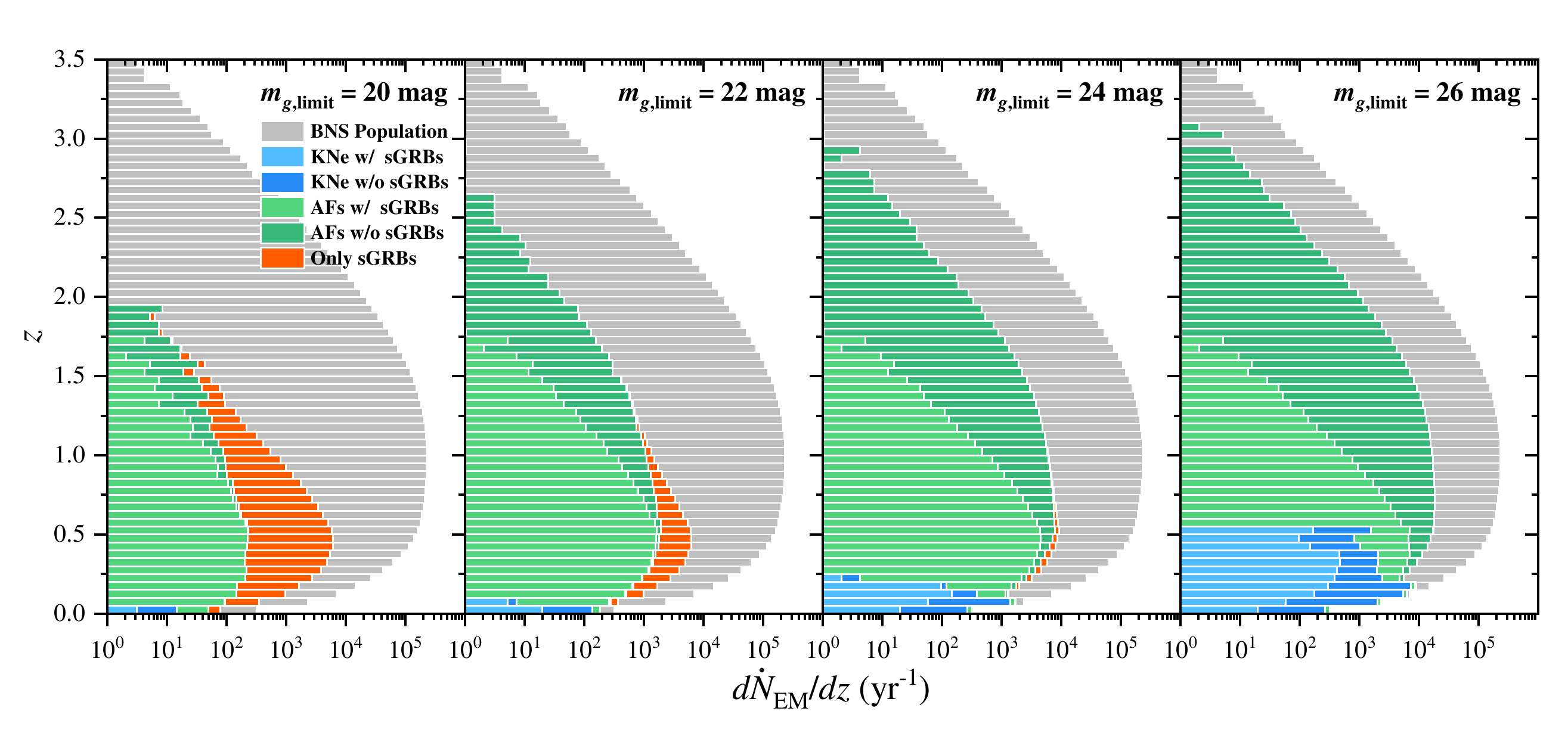}
    \caption{{{Redshift distributions of the detectable EM signals via the serendipitous searches for a $g$-band limiting magnitude of $m_{g,{\rm limit}} =20,\,22,\,24,\,{\rm and}\ 26\,{\rm mag}$ (from left to right panels). Gray histograms are the redshift distributions of the simulated cosmological BNS population. The light blue, dark blue, light green, dark green, and orange histograms represent the redshift distributions for the delectable samples of kilonovae w/ sGRBs, kilonovae w/o sGRBs, afterglows w/ sGRBs, afterglows w/o sGRBs, and  sGRBs only, respectively. The bin width of the histograms is set as $\Delta = 0.05$.}}}
    \label{fig:Population_Properties}
\end{figure*}

\begin{figure*}[htpb]
    \centering
    \includegraphics[width = 0.99\linewidth , trim = 8 10 10 40, clip]{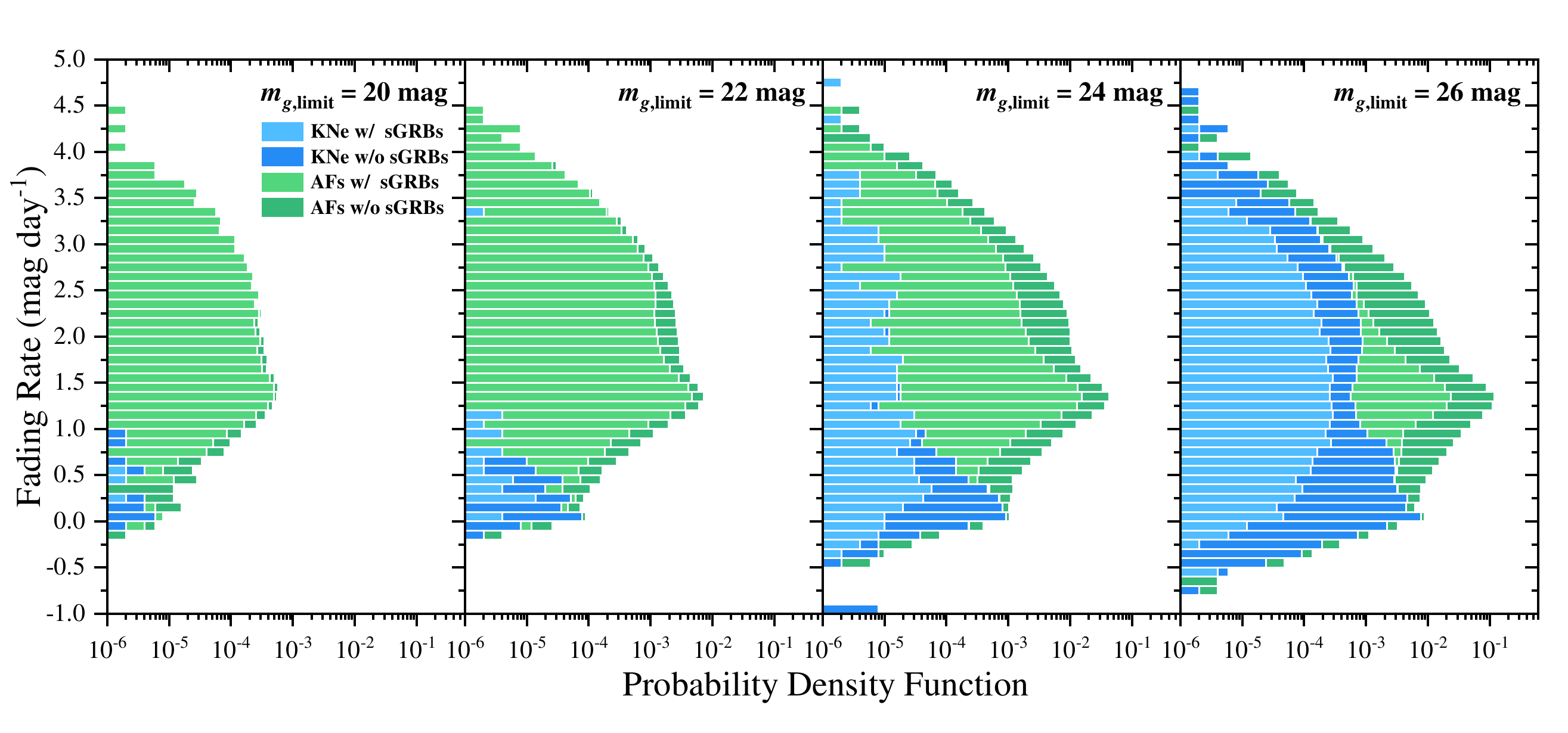}
    \caption{Similar to Figure \ref{fig:Population_Properties}, but for the $g$-band fading rates of the detectable EM signals. The bin width of the histograms is set as $\Delta = 0.1\,{\rm mag}\,{\rm day}^{-1}$.}
    \label{fig:Decaying_Rate}
\end{figure*}

{{By adopting an optimal serendipitous search strategy, i.e., an one-day cadence strategy, we show the redshift distributions of the detectable EM signals for a $g$-band limiting magnitude of $m_{g,{\rm limit}} =20,\,22,\,24,\,{\rm and}\ 26\,{\rm mag}$ in Figure \ref{fig:Population_Properties}. As for each detectable EM signal, we randomly simulate the detection epochs a thousand times and calculate the median difference value between these detection epochs as the fading rate. Figure \ref{fig:Decaying_Rate} shows the distributions of the fading rate for detectable EM signals. For the same search depth of each filter, there is no significant difference between searching in different bands. It is important to note that we collect all EM events whose $t_{\rm cad} \geq 1\,{\rm day}$ when we calculate the redshift distributions of the detectable EM signals, so that the distributions shown in Figure \ref{fig:Population_Properties} do not consider their detection probabilities. }}

{{For a limiting magnitude of $m_{\rm limit} = 20\,{\rm mag}$, the most likely EM counterpart of BNS mergers to be detected is individual sGRB emission. Due to this relatively shallow search depth, afterglow emissions associated with these individual sGRBs could have only at most one recorded epoch. In this case, it may be hard to establish the link between the sGRBs and the associated afterglows by the optically serendipitous searches. Detectable afterglow emissions should be much more easier to be discovered than kilonova emissions. However, these optical afterglows should be always associated with sGRB emissions. The most probable detectable redshift for these individual sGRBs and GRB-associated afterglows is $z\sim0.5$, which is consistent with the observations \citep[e.g.,][]{fong2015}. Some detectable orphan afterglows with much lower detection rate would take place at a range of $z\gtrsim0.75$. In our simulations, the largest distance of the detectable kilonovae is $z_{\rm max} \sim 0.02$ $(D_{\rm L,max}\sim 80\,{\rm Mpc})$. Most of ($\sim80\%-90\%$) these detectable kilonovae are expected to be discovered individually without the detections of accompanied sGRB emissions. }}

{{The improvement of the search depth would lead to a proportionate decrease in the individual detectable sGRB events. {{If $m_{\rm limit}\gtrsim22\,{\rm mag}$, more near-on-axis orphan afterglows and nearby off-axis orphan afterglows can be discovered, which would become the primary detectable EM counterparts of BNS mergers.}} For a limiting magnitude of $m_{\rm limit} \gtrsim 24\,{\rm mag}$, one can always find the associated afterglow and kilonova emissions after the sGRB triggers via the optically serendipitous searches. Due to the limited instrument sensitivity of {{$\gamma$-ray telescopes}}, the largest distance of the sGRB-associated afterglows and kilonovae is $z_{\rm max} \sim 1.75$. {{We note that this simulated largest distance of sGRB triggers is obtained by adopting an effective sensitivity limit of Fermi-GBM and GECAM. A few Swift sGRBs were found to have photometric redshifts of $z\gtrsim2$ presented by \cite{nugent2022}, because Swift-BAT has a lower sensitivity compared with Fermi-GBM and GECAM, and hence a deeper detection depth. }} With the increase of the search depth, kilonovae would play a leading role of nearby detectable EM counterparts. For a limiting magnitude of $m_{g,{\rm limit}} =22,\,24,\,{\rm and}\ 26\,{\rm mag}$, the median (largest) distances of these detectable kilonovae are $z = 0.04,\,0.1,\,{\rm and}\ 0.25$ ($z_{\rm max} = 0.06,\,0.21,\,{\rm and}\ 0.55$). Search depth has little effect on the ratio between detectable kilonovae w/ sGRBs and kilonovae w/o sGRBs.}}

{{The fading rates of the detectable afterglows always peak at $\sim1.3\,{\rm mag}\,{\rm day}^{-1}$, which have a wide distribution between $\sim -0.5\,{\rm day}^{-1}$ and $\sim 4.5\,{\rm mag}\,{\rm day}^{-1}$. Comparing with afterglows, kilonova-dominated events have more slow-evolving lightcurves. Their fading rates peak at $\sim 0 - 0.1\,{\rm mag}\,{\rm day}^{-1}$. For a limiting magnitude of $m_{\rm limit}\lesssim22\,{\rm mag}$, the fading rates of kilonovae locate in a range from $-0.25\,{\rm day}^{-1}$ to $\sim1\,{\rm mag}\,{\rm day}^{-1}$. As shown in Figure \ref{fig:Decaying_Rate}, by adopting a limiting magnitude of $m_{\rm limit}\gtrsim24\,{\rm mag}$ ($m_{\rm limit}\gtrsim26\,{\rm mag}$), some fast-evolving sGRB-associated kilonovae (kilonovae w/ sGRBs and kilonovae w/o sGRBs) with a fading rate of $\gtrsim 1\,{\rm mag}\,{\rm day}^{-1}$ can be discovered. For these fast-evolving kilonova events, their early-stage observations would be contributed by the associated afterglows while the kilonova emissions would lead to the late-stage observations. }}


\section{Detectability for Target-of-opportunity Observations of GW Triggers} \label{sec:too}

\subsection{GW Detectability Method \label{sec:GWDetectionMethod}}

It is expected that two Advanced LIGO detectors (H1 and L1) in the USA \citep{harry2010,aasi2015}, Advanced Virgo detector (V1) in Europe \citep{acernese2015}, and KAGRA detector (K1) in Japan \citep{aso2013,akutsu2019} will start the fourth observation run (O4) together in 2023. {{The network composed of these 2nd generation detectors is referred to as the “HLV era” in the following. Here, the sensitives of H1, L1 and V1 in the HLV era are adopted as their respective design sensitivities \citep{abbott2020prospects} since their sensitivities are dynamic and change over time\footnote{{{We show differences between design sensitivity curves we use and latest sensitivity curves released on April 6th, 2022 in Figure \ref{fig:ASD_Comparsion} of Appendix \ref{app:ASD}. Based on these latest sensitivity curves, our simulations of GW detection rate in O4 might be slightly overestimated.}}}, while K1 is ignored in our simulations in view of that K1 will work to improve most of time in O4\footnote{\url{https://www.ligo.caltech.edu/news/ligo20220617}}.}} The 2nd generation detectors would finish their upgrade to 2.5th generation detectors in $\sim 2025$. The subsequent upgrade of Advanced LIGO, Advanced Virgo, and KAGRA are called Advanced LIGO Plus \citep[A+;][]{miller2015}, Advanced Virgo Plus \citep[AdV+;][]{abbott2020prospects}, and KAGRA+ \citep{michimura2020}. Hereafter, we refer to the era during which these four detectors upgrade to 2.5th generation detectors as the “PlusNetwork era”. After $\sim2030$, the 3rd generation GW detectors are expected to start their observation. The currently proposed 3rd generation detector plans include LIGO Voyager \citep{adhikari2020} as a possible upgrade upon LIGO A+ (strictly speaking, it's more like quasi-3rd generation. However, since its sensitivity is much higher than that of the 2.5th generation detectors, for the convenience of discussion, we classify it as 3G), ET in Europe \citep{punturo2010a,punturo2010b,maggiore2020}, and CE in the USA \citep{reitze2019}. Due to the as-yet undetermined locations of ET and CE, we directly place ET at the current Virgo detector position and two CE detectors at the current H1 and L1 positions, according to the convention \citep{vitale2017,vitale2018}.

For each BNS system, we randomly simulate the masses of individual NSs based on the observationally derived mass distribution of Galactic BNS systems, i.e., a normal distribution $M_{\rm NS}/M_\odot\sim\mathcal{N}(1.32, 0.11^2)$ \citep{lattimer2012,kiziltan2013}. The NS equation of state (EoS) DD2 \citep{typel2010}, which is one of the stiffest EoS allowed by present constraints \citep[e.g.,][]{gao2016,abbott2019properties}, is adopted. With known $M_{\rm NS}$, $z$, and EoS, we use the \texttt{IMRPhenomPv2\_NRTidalv2} \citep{dietrich2019} waveform model to simulate the GW waveform in the geocentric coordinate system, and then project it to different detectors to obtain the detector-frame strain signal. The optimal signal-to-noise ratio (S/N) can be obtained by
\begin{equation}
    \rho_{\rm opt}^2 = 4\int_{f_{\rm min}}^{f_{\rm max}} \frac{|\tilde{h}(f)|^2}{S_n(f)}df
     = \int_{f_{\rm min}}^{f_{\rm max}} \frac{(2|\tilde{h}(f)|\sqrt{f})^2}{S_n(f)}d\,{\rm ln}(f),
\end{equation}
where $f$ is the frequency, $\tilde{h}(f)$ is the strain signal in the frequency domain, and $S_n(f)$ is the one-sided power spectral density of the GW detector which is square of the amplitude spectral density (ASD). The ASD for each detector is shown in the Appendix \ref{app:ASD}. We set the maximum frequency $f_{\rm max}$ as $2048\,{\rm Hz}$. The low frequency cutoff $f_{\rm min}$ is set to $20\,{\rm Hz}$ for O3, $10\,{\rm Hz}$ for all the 2nd, 2.5th generation detectors \citep{miller2015} and LIGO Voyager \citep{adhikari2020}, $5\,{\rm Hz}$ for CE \citep{reitze2019}, and $1\,{\rm Hz}$ for ET \citep{punturo2010a}. We use the optimal S/N to approximate the matched filtering S/N of the GW signal detected by each detector, and then calculate the network S/N of the entire detector network, i.e., the root sum squared of the S/N of all detectors. In each GW era, when the S/N for a single detector is greater than the threshold of 8 and the network S/N is greater than 12, we expect that the corresponding GW signal is detected. 

\begin{table}[t] 
\centering
\caption{{{O3 duty cycle \label{tab:Duty}}}}
\begin{tabular}{cc}
\hline\hline
{Online Detector} & {$P_{\rm duty}$} \\ \hline
HLV & $46.75\%$ \\
HL  & $14.80\%$ \\
HV  & $9.68\%$  \\
LV  & $11.8\%$  \\
H   & $3.06\%$  \\
L   & $2.94\%$  \\
V   & $7.59\%$  \\
None & $3.35\%$ \\
\hline
\end{tabular}
\end{table}

{{We consider the exact duty cycle of O3 (see Table \ref{tab:Duty}), calculated following the timeline released from LVC\footnote{\url{https://www.gw-openscience.org/O3/index/}}, to simulate the GW observations of BNS mergers in O3 and O4. In view of significant technology upgrades for GW detections and more detectors that will join GW campaigns, the duty cycle in the 2.5th and 3rd generation detector networks could be highly uncertain. Thus, we only calculate their best cases, i.e., ``all detectors in the corresponding era have reached the design sensitivity and work normally" as the optimal situation. During the 2.5th generation detector network, the best case is that A+, AdV+, and KAGRA+ all work normally, which we abbreviate as “PlusNetwork”. LIGO Voyager is separately discussed. Furthermore, ``ET\&CE” represents the best case of the 3rd generation era.}}

{{We need to localize the BNS through GW signals for EM follow-ups. Since we need to calculate a large number of simulated signals, we use Fisher Information Matrix \citep[FIM;][]{cutler1994gravitational} to approximate the localization area estimated by the more computationally expensive Bayesian method \citep{thrane2019introduction}. The FIM is based on the Linear Signal Approximation \citep[LSA;][]{cutler1994gravitational}, and {{uses a Gaussian distribution}} to approximate the posterior distribution of the parameters. This assumption requires the signal to have a sufficiently high S/N, so we only calculate the GW localization for the signal whose network S/N meets the detection threshold. The FIM of the detector network is a linear summation of the FIM of the individual detector in that network}}
\begin{equation}
\label{equ:fim_net}
\boldsymbol{\Gamma}_{i j}=\sum_{k}\left\langle \partial_{i} h \mid \partial_{j} h \right\rangle_{k},
\end{equation}
{{where the bracket means the inner product}}
\begin{equation}
\label{equ:inner_product}
\left\langle a \mid b\right\rangle=4 \Re \int_{f_{\text {min}}}^{f_{\text {max}}} \frac{\tilde{a}(f) \tilde{b}^{*}(f)}{S_{n}(f)} \mathrm{~d} f,
\end{equation}
{{$k$ is the index of the detector in that network, $\partial_{i} h$ or $\partial_{i} h$ refers to the partial differentiation of the detector-frame signal in the frequency domain with respect to a certain parameter. In our FIM calculation, the parameters are chosen from the detector-frame chirp mass $\mathcal{M}$, the symmetric mass ratio $\eta$, the luminosity distance $D_{L}$, the coalescence time $t_{c}$, the coalescence phase $\phi_{c}$, the inclination angle $l$, the polarization angle $\psi$, the right ascension $\theta$, the declination $\phi$, and the tidal deformation parameters $\tilde{\Lambda}$ and $\delta \tilde{\Lambda}$. Note that, for the 3rd generation detector network, we also take the Earth's rotation into account \citep{2022ApJ...926..158L}. In order to reduce the matrix singularity issue, we don't take partial differentiation of the tidal parameters for the cases before the 3rd generation.}}

{{For high-S/N signals, the inverse of the FIM is less or equal to the covariance matrix of parameters, the so-called ``Cramer-Rao lower bound''}}
\begin{equation}
\label{equ:covariance_matrix}
\operatorname{cov}\left(i, j\right) \geq\left(\boldsymbol{\Gamma}^{-1}\right)_{i j},
\end{equation}
{{for a specific parameter, we can use the square root of the corresponding diagonal element in the inverse of the FIM as the bias. In our case, we care about $\Delta \cos \theta$ and $\Delta \phi$. Then we can get the sky localization area}} \citep{2004PhRvD..69h2005B}
\begin{equation}
\label{equ:sky_area}
\Omega_{\rm GW}=2 \pi \sqrt{(\Delta \cos \theta \Delta \phi)^{2}-\langle\Delta \cos \theta \Delta \phi\rangle^{2}},
\end{equation}
{{we use 90\% confidence of this area hereafter.}}

\subsection{GW Detections and EM Follow-ups in the 2nd Generation Era}

\subsubsection{GW Detection Rate, Detectable Distance, and Sky Localization}

\begin{table*}[htpb] 
\centering
\caption{{{GW Detection Results}}} \label{tab:GWDetection}
\begin{tabular}{ccccccc}
\hline\hline
\multirow{2}{*}{Case}   & \multirow{2}{*}{Era} & $\dot{N}_{\rm GW}/{\rm yr}^{-1}$ & $z$& $z_{\rm max}$ & \multicolumn{2}{c}{$\log_{10}(\Omega_{\rm GW}/{\rm deg}^2) = a\times \log_{10}(z) + b$}\\ \cline{6-7} 
&  & $(\dot{N}_{\rm GW}/\dot{N}_{\rm BNS})$ & $(D_{\rm L}/100\,{\rm Mpc})$ & $(D_{\rm L,max}/100\,{\rm Mpc})$  & $a$ & $b$  \\ \hline
\multirow{2}{*}{HLV (O3)}& \multirow{2}{*}{2nd}   &  $2.4^{+3.6}_{-1.8}$  & $0.025^{+0.025}_{-0.013}$ & $0.062$
 & \multirow{2}{*}{$-$} & \multirow{2}{*}{$-$}\\
&& $(0.001\%)$ & $(1.1^{+1.2}_{-0.5})$ & $(2.9)$\\
\multirow{2}{*}{HLV (O4)} & \multirow{2}{*}{2nd} & $11^{+17}_{-8}$ & $0.040^{+0.025}_{-0.025}$ & $0.084$ &\multirow{2}{*}{$1.93$} & \multirow{2}{*}{$4.02^{+0.85}_{-0.42}$}     \\
& & $(0.004\%)$ & $(1.8^{1.2}_{1.1})$ & $(4.0)$\\ \hline
\multirow{2}{*}{PlusNetwork}  & \multirow{2}{*}{2.5th} & $210^{+320}_{-160}$  & $0.099^{+0.050}_{-0.055}$  & $0.190$  & \multirow{2}{*}{$1.87$} & \multirow{2}{*}{$2.85^{+0.45}_{-0.55}$}  \\
&  & $(0.08\%)$     & $(4.7^{+2.6}_{-2.7})$& $(9.5)$\\ \hline
\multirow{2}{*}{LIGO Voyager} & \multirow{2}{*}{3rd}   & $1.8^{+2.8}_{-1.4}\times10^3$    & $0.22^{+0.12}_{-0.13}$     & $0.43$ & \multirow{2}{*}{$-$} & \multirow{2}{*}{$-$}  \\
&  & $(0.73\%)$     & $(11.0^{+7.4}_{-6.7})$     & $(24.3)$  \\
\multirow{2}{*}{ET\&CE}  & \multirow{2}{*}{3rd} & $2.4^{+3.6}_{-1.8}\times10^5$    & $0.97^{+0.71}_{-0.57}$     & $3.77$ & \multirow{2}{*}{$2.00$} & \multirow{2}{*}{$0.85^{+0.69}_{-0.46}$} \\
&  & $(90.7\%)$     & $(65^{+64}_{-43})$   & $(343)$   \\ \hline
\end{tabular}
\tablecomments{GW detection rates and luminosity distance distributions for detectable GWs in the 2nd generation era are simulated by adopting an exact duty cycle labeled in Table \ref{tab:Duty}, while GW detection results in the 2.5th and 3rd generation eras are obtained with consideration of ideal operation conditions. The columns are [1] the case of different generation eras; [2] the generation of GW detectors; [3] median GW detection rates with consideration of $90\%$ interval by adopting the local event rate density of $\dot{\rho}_{0,{\rm BNS}} = 320^{+490}_{-240}\,{\rm Gpc}^{-3}\,{\rm yr}^{-1}$ \citep{abbott2021population}, while the numbers in brackets are the corresponding detectable proportions of the number of BNS mergers per year in the universe ($\dot{N}_{\rm BNS}$); [4] median detectable redshifts and detectable luminosity distances with consideration of $90\%$ intervals; [5] maximum detectable reshifts and detectable luminosity distances; {{[6] GW sky localizations as function of $z$, where $a$ and $b$ are fitting parameters.}}}
\end{table*}

\begin{figure}
   \centering
   \includegraphics[width = 0.99\linewidth , trim = 75 30 70 60, clip]{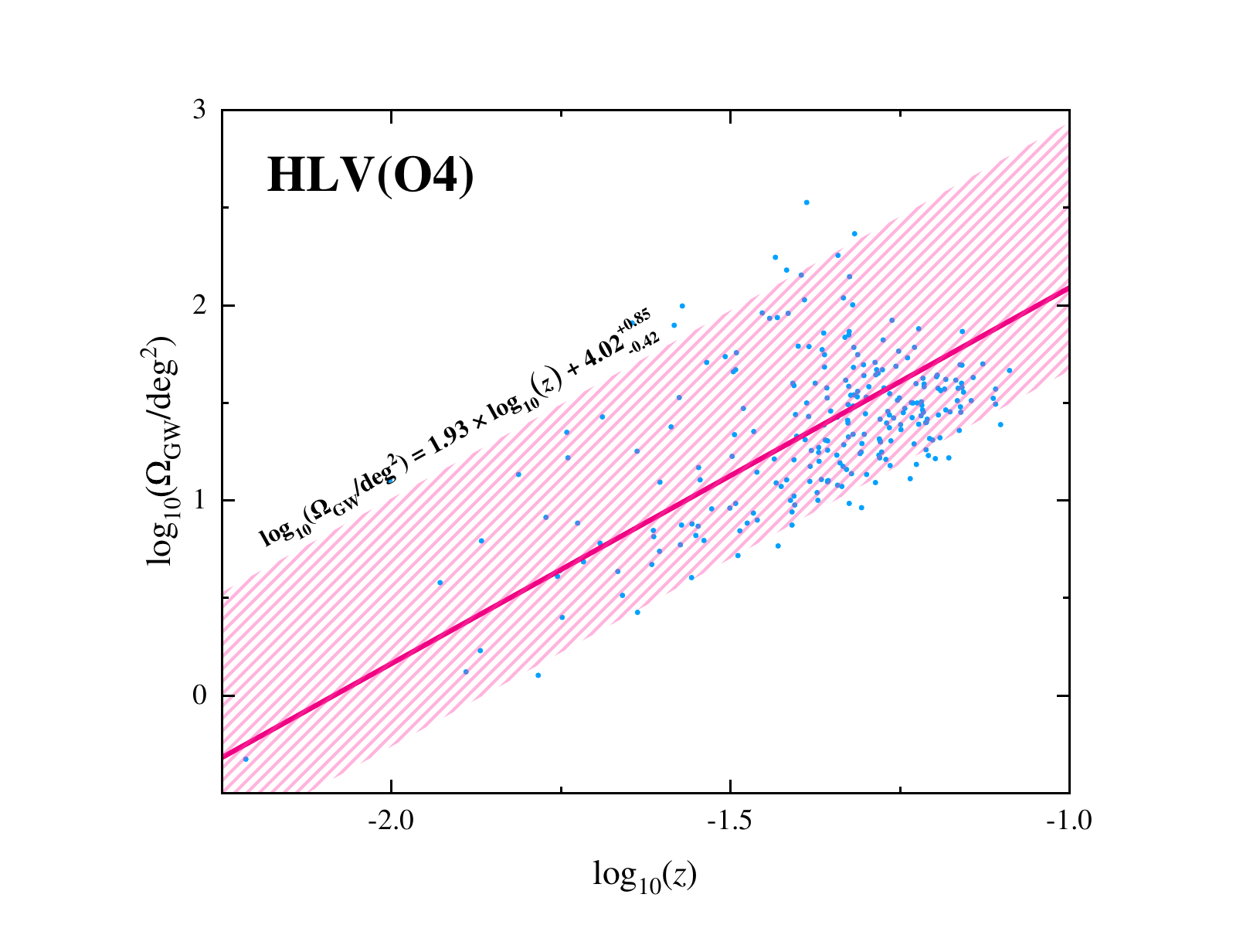}
   \caption{{{$90\%$ confidence of GW sky localization vs. redshift for BNS mergers detected at the HLV (O4) era when all three GW detectors are online at the same time.}}  }
    \label{fig:GWOmega}
\end{figure}

{{By considering an exact duty cycle shown in Table \ref{tab:Duty}, the simulated GW detection results of O3 and O4 are summarized in Table \ref{tab:GWDetection}. We check that the GW detection rate in O3 should be $\sim2.4^{+3.6}_{-1.8}\,{\rm yr}^{-1}$, which is consistent with the observations of LVC \citep{abbott2020prospects,abbott2021gwtc2}. The median detectable luminosity distance is $\sim110\,{\rm Mpc}$, nearly approximate to the observed distance of GW190425 \citep{abbott2020gw190425}. In the HLV (O4) era, we predict that one can detect $\sim11\,{\rm yr}^{-1}$ BNS GW events with a median detectable distance at $z \sim 0.040$ and a horizon at $z_{\rm max} \sim 0.084$.}}

\begin{figure}
   \centering
   \includegraphics[width = 0.99\linewidth , trim = 75 30 70 60, clip]{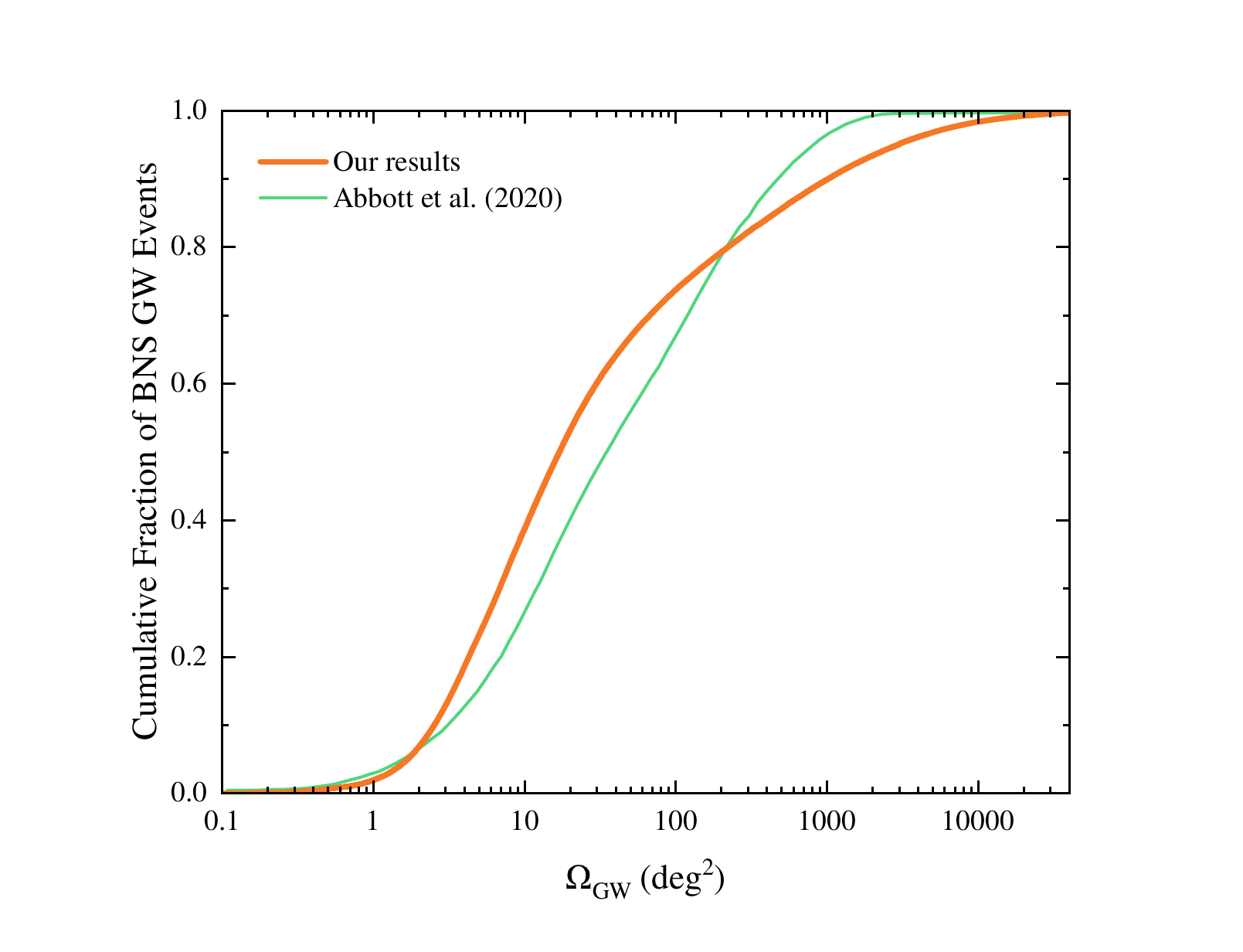}
   \caption{{{The cumulative fractions of BNS GW events with sky localization area during O4 smaller than the abscissa value. Our simulation results are marked as thick orange line, while model distribution from \cite{abbott2020prospects} is shown as thin green.}} }
    \label{fig:GWOmega_CDF}
\end{figure}

{{We simulate the sky localization area ($\Omega_{\rm GW}$) for detectable BNS GW events when two or three detectors are online simultaneously during O4. Since the localization for GW mainly rely on the time delay between different detectors, so one 2nd generation GW detector is impossible to localize GW signals. We find that the relationship between the sky localization and redshift for BNS mergers detected in O4 with the H1, L1 and VIRGO network can be well explained by a log-linear trend (Figure \ref{fig:GWOmega}), while the log-linear relationships with only two GW detectors network are not obvious due to their limited detections. In Figure \ref{fig:GWOmega_CDF}, the median sky localization area is expected to be $\sim10\,{\rm deg}^2$ for detectable BNS GW events. We also collected the cumulative fraction of the sky localization from \cite{abbott2020prospects}. Comparing with our simulation, they considered the operation of K1 in O4 and used \texttt{BAYESTAR} \citep{singer2016} code to perform sky localization of BNS GW events. A duty cycle of $70\%$ for each detector uncorrelated with the other detectors, which is slightly different with our simulations, was adopted by \cite{abbott2020prospects}. However, our simulation result for the sky localization in O4 is nearly consistent with that shown in \cite{abbott2020prospects}. }}

\subsubsection{{{EM Detectability in O4}}}

{{Based on the BNS GW detection results during O4, we now estimate the EM detection rates of detectable BNS GW events for ZTF, SiTian, Mephisto, WFST, and LSST. For the serendipitous observations, the event would appear in an arbitrary position of the celestial sphere due to the lack of the ToO alerts. For the GW-triggered ToO observations, the survey project would just need to cover the sky localization of GW events in search of their associated EM counterparts. Thus, one can replace $\Omega_{\rm sph}$ with $\Omega_{\rm GW}$ in Equation (\ref{equ:EMdetection_rate}) to estimate the EM detection rate for the ToO observations, i.e.,}}
\begin{equation}
\begin{split}
\label{equ:GWEMdetection_rate}
    \dot{N}_{\rm EM} \approx &\frac{\dot{N}_{\rm BNS}}{n_{\rm sim}} \cdot \frac{\Omega_{\rm cov}}{\Omega_{\rm sph}}  \\ \times&\sum_{i = 1}^{n_{\rm GW}}\sum_{j}\frac{\Omega_{\rm FoV} P_{{\rm duty},j}\min(t_{{\rm cad},ij},\Delta t_{ij})}{\max(\Omega_{\rm FoV},\Omega_{{\rm GW},j})(n_{{\rm exp},ij}t_{{\rm exp},ij} + t_{\rm oth})},
\end{split}
\end{equation}
{{where $j = \{\rm HL, HV, LV, HLV\}$ and $n_{\rm GW}$ represents detectable BNS GW events. Here, we adopt $t_{\rm exp} = 300\,{\rm s}$ to make GW-triggered follow-up observations. Thus, the cadence time $t_{\rm cad}$ for each event is related to $t_{\rm exp}$ and $\Omega_{\rm GW}$ for each event. The judgement condition for the detection of the kilonova and/or afterglow by a follow-up search after GW triggers is required to be \emph{two different exposure filters have at least two detection epochs}.}}

{{Since SiTian will not operate during O4, we only show our simulated detection rates for ZTF, Mephisto, WFST and LSST at this era. Based on the technical informations of the survey projects listed in Table \ref{tab:SurveyProject}, our simulation results show that ZTF/SiTian/Mephisto/WFST/LSST can detect $\sim5/4/3/3$ kilonovae ($\sim1/1/1/1$ afterglows) in O4, respectively. $\sim5\%$ kilonovae and $\sim90\%$ afterglows after GW triggers are expected to be associated with the detections of sGRBs.}}

\subsection{GW Detections and EM Follow-ups in the 2.5th and 3rd Generation Eras}

\subsubsection{GW Detection Rate, Detectable Distance, and Sky Localization}  \label{sec:GWResults2}

\begin{figure*}[htpb]
    \centering
    \includegraphics[width = 0.99\linewidth , trim = 95 280 135 50, clip]{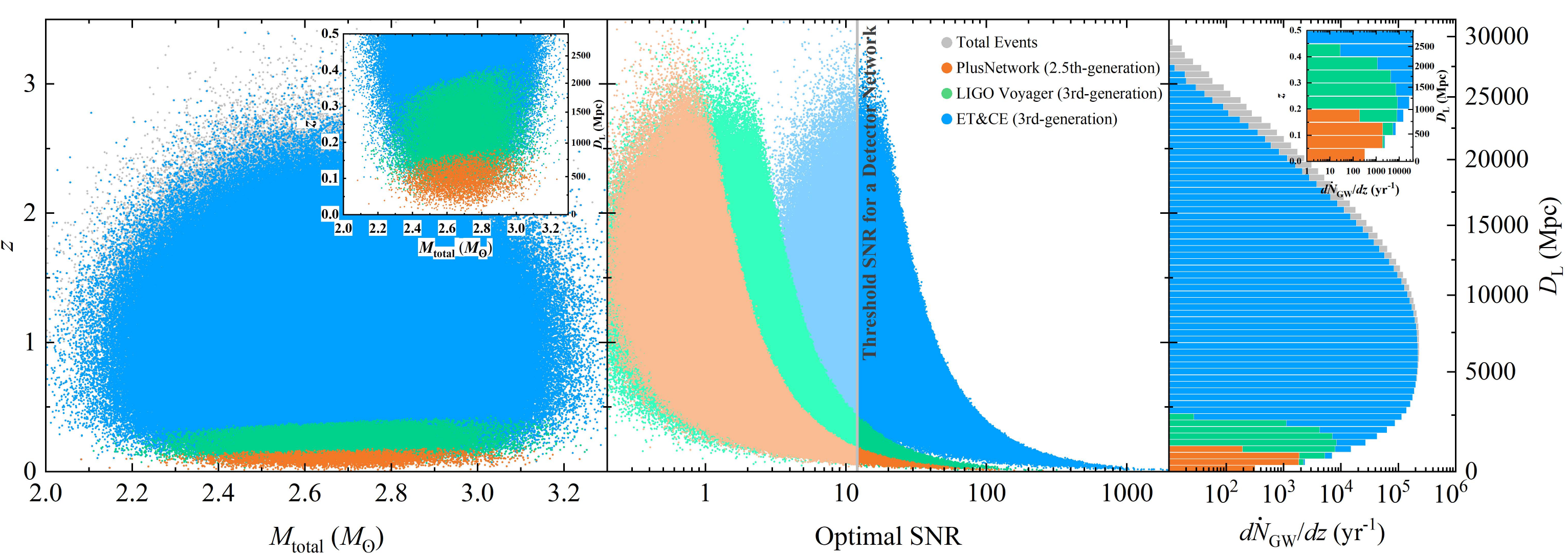}
    \caption{{{Detectability of BNS mergers by various detector networks in different GW detection eras. The left panels show the signals that can be detected by different detectors and detector networks. The orange, green, and blue dots are the GW signals detectable at PlusNetwork, LIGO Voyager, and ET\&CE network eras, respectively. The gray dots in the background represent undetectable signals. The small panels in the upper right corner are enlarged images of the low-redshift area. The middle panels show the distributions of all simulated signals on the “optimal S/N–redshift” plane, with detection thresholds for a detector network (i.e., S/N = 12; light gray line). To the right of the thresholds are the GW signals that can be detected. The right panels show the distributions of BNS detection rates with redshift, and the insets are zoom-in pictures in the low-redshift region.}}}
    \label{fig:GW2}
\end{figure*}

{{We summarize all our simulated GW detection results of 2.5th and 3rd generation eras in Table \ref{tab:GWDetection}. The total mass, S/N, and redshift for detectable GW signals in different eras are shown in Figure \ref{fig:GW2}. For the 2.5th generation GW detector network, the optimal detection rate is $\sim210\,{\rm yr}^{-1}$. The GW detection distance would be doubled compared with the detection distance in O4, i.e., the horizon can reach $z_{\rm max}\sim0.2$. For the LIGO Voyager in the 3rd generation era, the optimal detection rate can be increased to $\sim 1,800\,{\rm yr}^{-1}$ and the detection distance would be twice compared with the last era, i.e., a horizon of $z_{\rm max} \sim 0.4$. However, these numbers are much smaller than those for the newly designed 3rd generation detectors. For the ET\&CE network, the optimal detection rate would be $\sim2.4\times10^{5}\,{\rm yr}^{-1}$ which would account for $\sim91\%$ of the total BNS GW events in the universe. The events detected by ET\&CE are mainly dominated by BNS mergers at $z \sim 1$, which is near the most probable redshift where BNS mergers occurred in the universe. The most remote detectable events by ET\&CE would be at $z_{\rm max} \sim 3.8$. Except for the ET\&CE era, the median distance of detectable GW events is always set at half of the horizon in each GW era.}}

\begin{figure}
   \centering
   \includegraphics[width = 0.99\linewidth , trim = 75 30 70 60, clip]{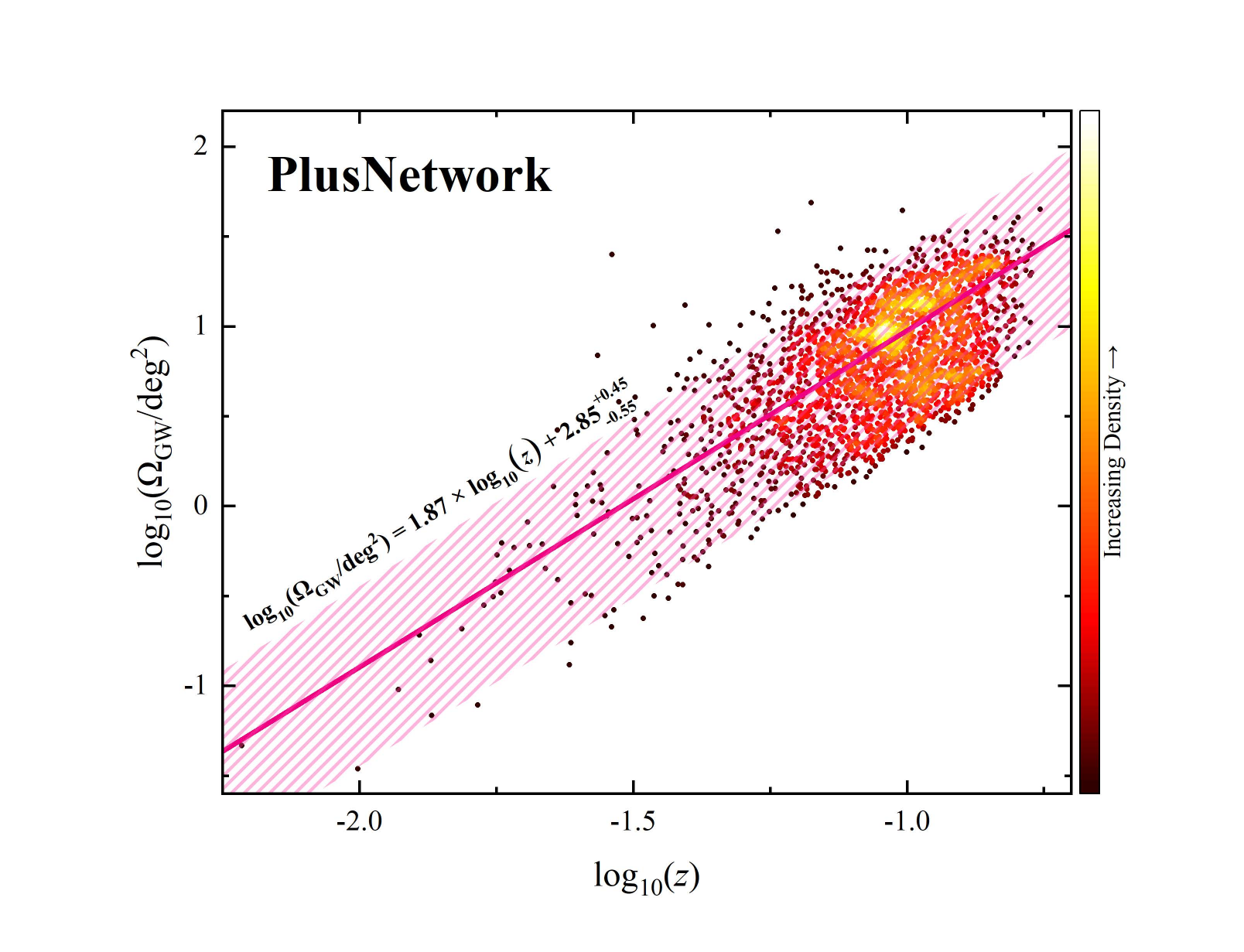}
   \includegraphics[width = 0.99\linewidth , trim = 75 30 70 60, clip]{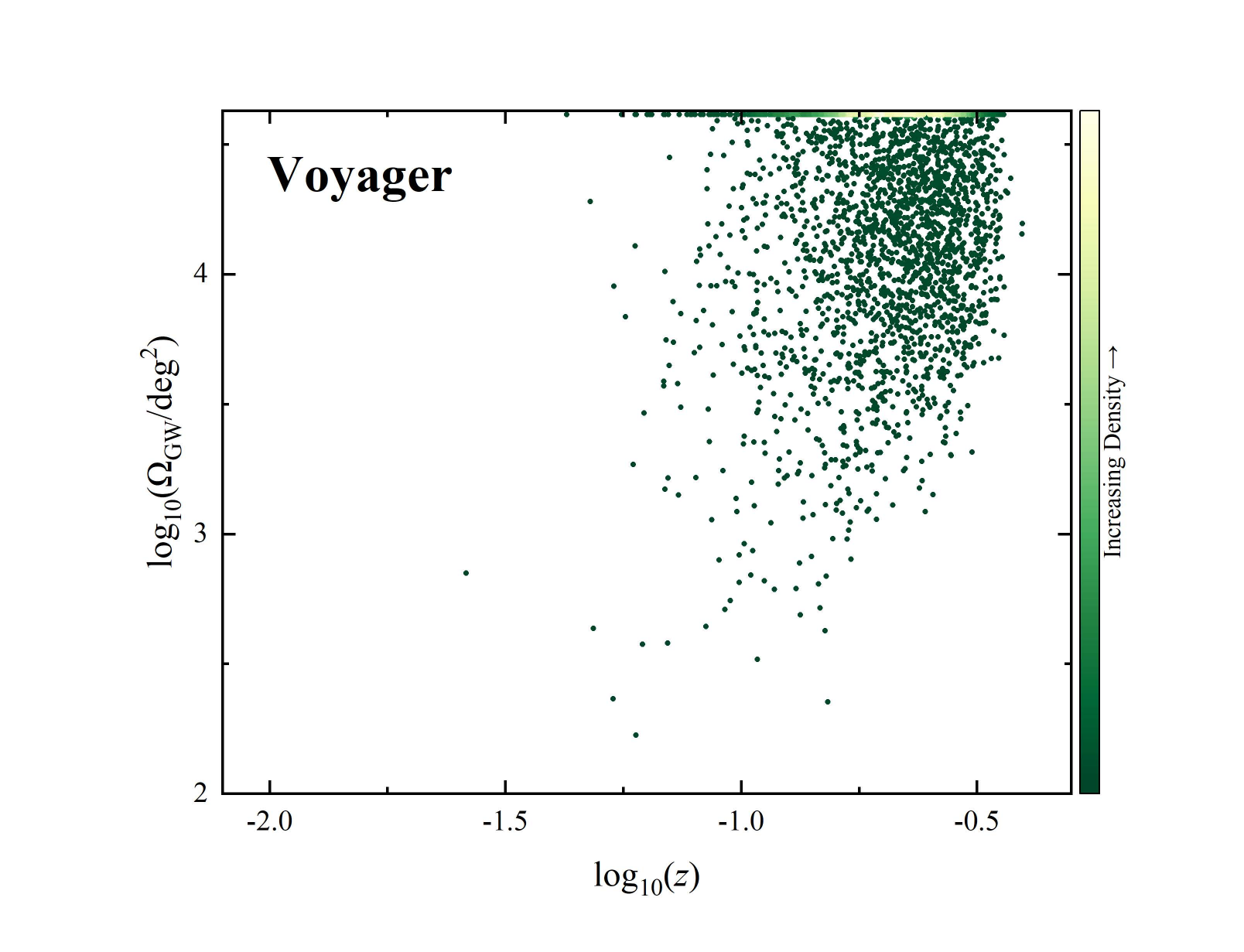}
   \includegraphics[width = 0.99\linewidth , trim = 75 30 70 60, clip]{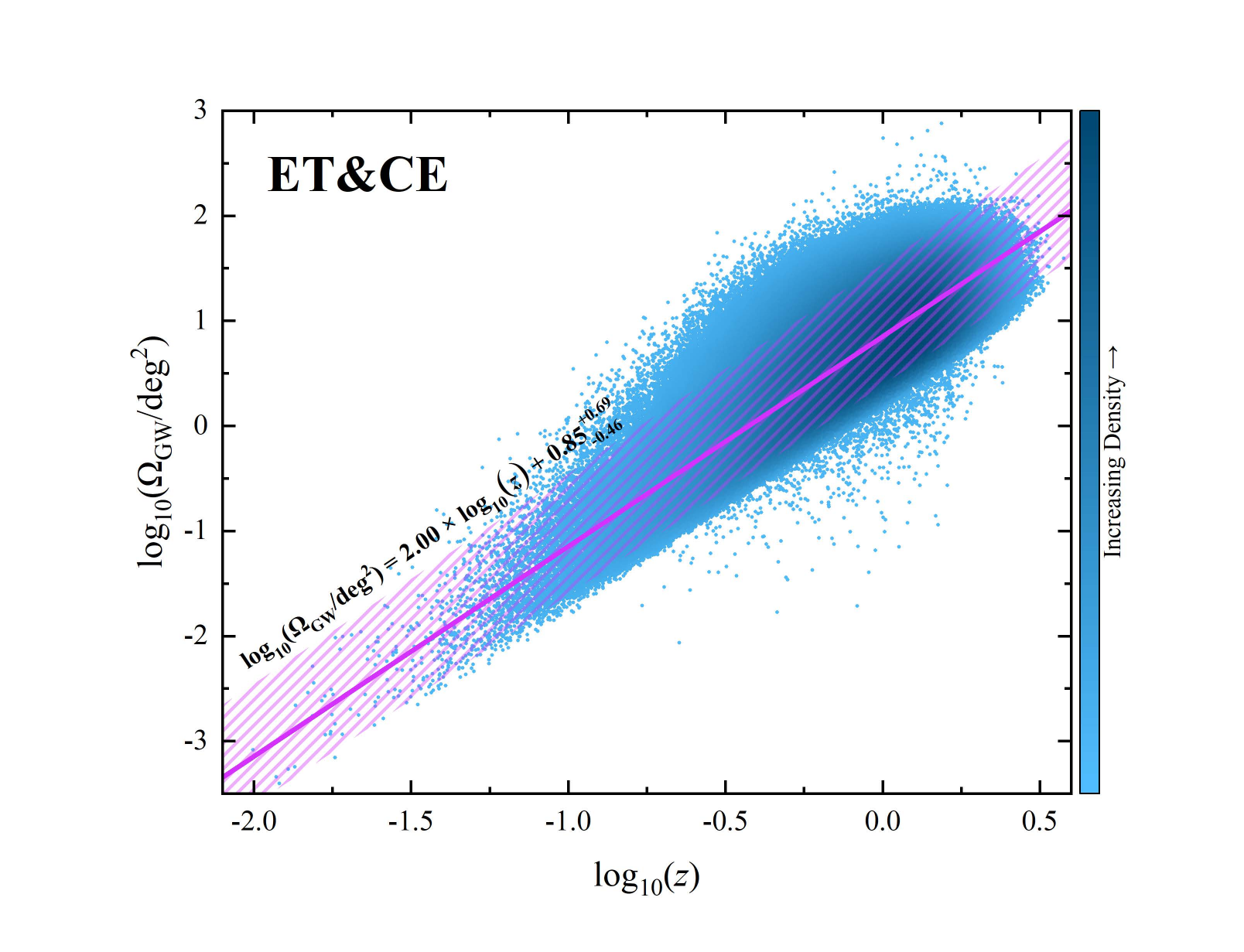}
   \caption{{{$90\%$ confidence of GW sky localization vs. redshift for BNS mergers detected at the PlusNetwork (top panel), LIGO Voyager (middle panel) and ET\&CE eras, respectively. The density (see the colorbar of each panel) for the points is calculated via the kernel density estimation. Solid line and shaded region in each panel represent the median of the GW sky localization and $90\%$ interval, respectively.  }}}
    \label{fig:GWOmega2}
\end{figure}

{{In Figure \ref{fig:GWOmega2}, the relationships between the $90\%$ credible area of GW sky localization and redshift for BNS mergers detected at the PlusNetwork, LIGO Voyager and ET\&CE eras can be represented by log-linear trends, similar to that of 2nd generation network. The fitting results of these log-linear trends are listed in Table \ref{tab:GWDetection}. In the PlusNetwork era, most of detectable BNS mergers will be localized to $\lesssim10\,{\rm deg}^2$. The network of one ET detector and two CE detectors can have a more remarkable capability to observe and localize BNS GW events. For BNS mergers that occurring at $z\lesssim0.2$, the GW sky localizations constrained in the ET\&CE era will be about two orders of magnitude lower than those constrained in the PlusNetwork era. The median localization for BNS mergers at $z\sim0.5$ ($z\sim1$) is shown to be $\sim1\,{\rm deg}^2$ ($\sim10\,{\rm deg}^2$). In these regimes, the present and future wide-field-of-view survey projects will be able to cover the sky localizations given by GW detections in a few pointings and achieve deep detection depths with relatively short exposure integration times. In view of that current GW operation plan during the LIGO Voyager era only includes two GW detectors, we find that the sky localizations will span from a few hundred square degrees to tens of thousands of square degrees. Thus, EM follow-ups might be very difficult if there are no more GW detectors join the campaign at the LIGO Voyager era.}}

\subsubsection{EM Detectability}

\begin{figure}
   \centering
    \includegraphics[width = 0.99\linewidth , trim = 65 55 95 60, clip]{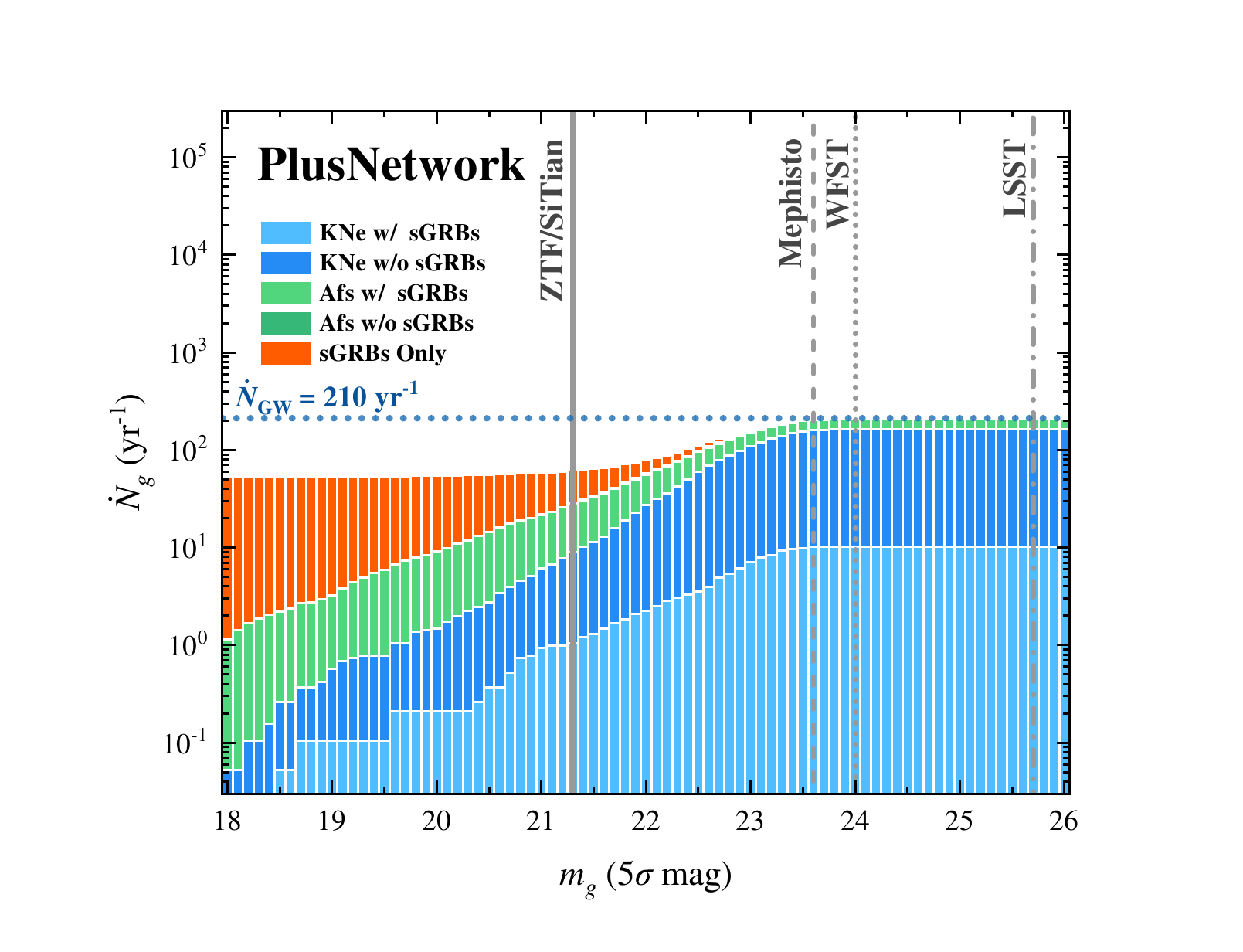}
    \includegraphics[width = 0.99\linewidth , trim = 65 55 95 60, clip]{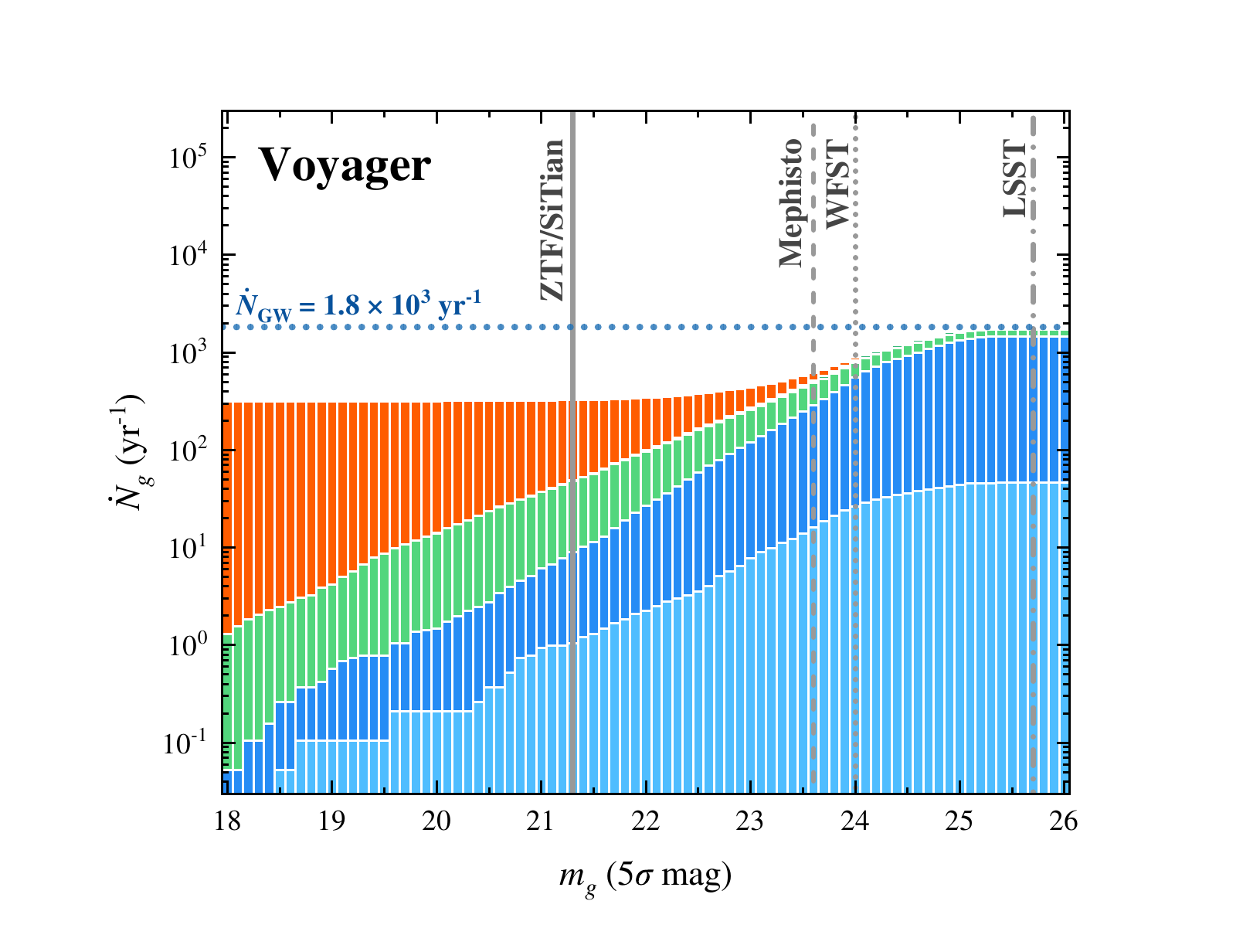}
    \includegraphics[width = 0.99\linewidth , trim = 65 30 95 60, clip]{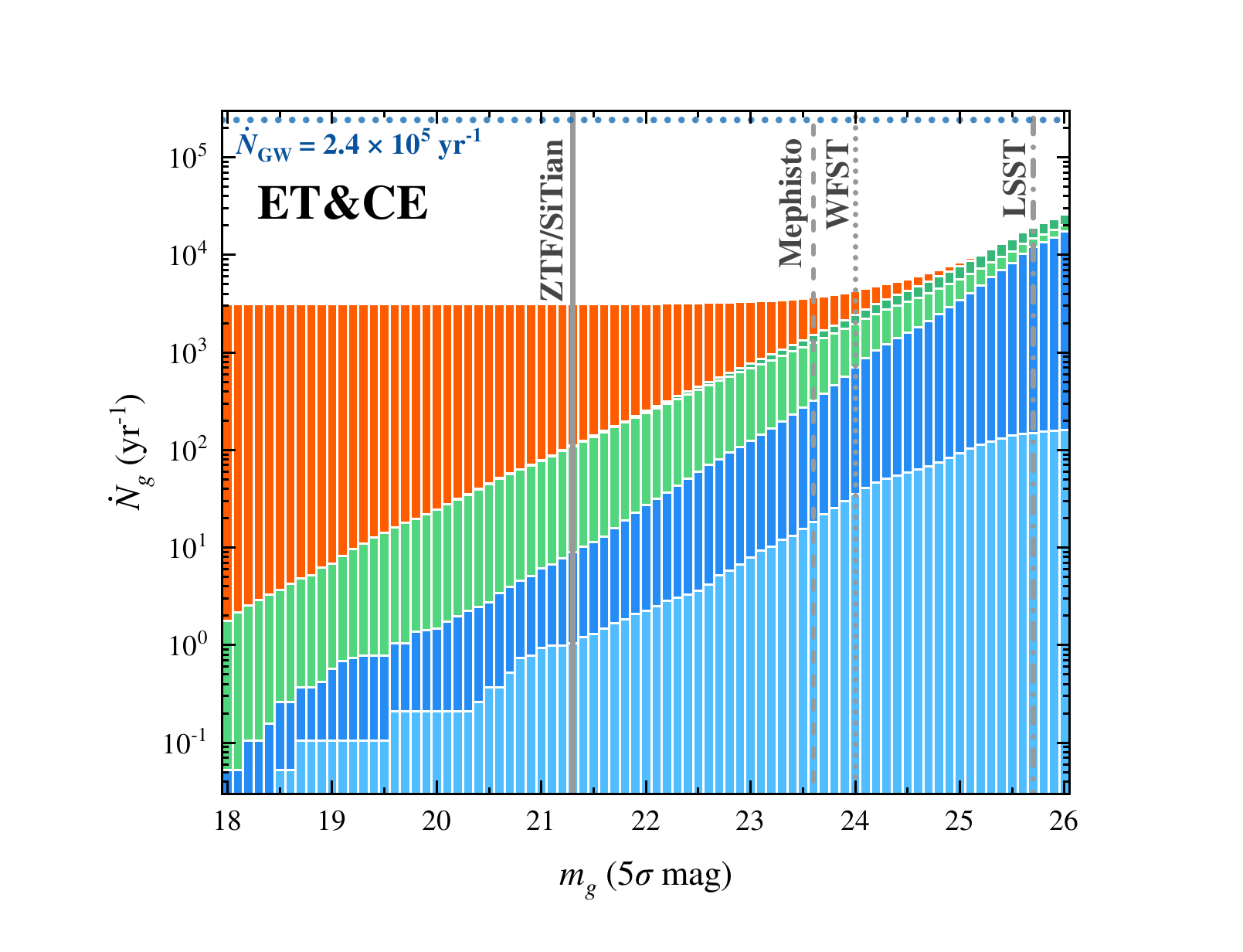}
    \caption{{$g$-band luminosity functions for the delectable samples of kilonovae w/ sGRBs (light blue histograms), kilonovae w/o sGRBs, (dark blue histograms), afterglows w/ sGRBs (light green histograms), afterglow w/o sGRBs (dark green histograms), and only sGRBs (orange histograms) as functions of $g$-band $5\sigma$ limiting magnitude during three GW detection eras, i.e., the PlusNetwork, LIGO Voyager, and ET\&CE eras. The gray solid, dashed, dotted, and dashed-dotted lines respectively represent the $r$-band $5\sigma$ limiting magnitude of ZTF/SiTian, Mephisto, WFST, and LSST, with $300\,{\rm s}$ exposure time. The dashed blue lines shows the GW detection rates in each GW era. The bin width of the histograms is set as $\Delta = 0.1\,{\rm mag}$.}}
    \label{fig:GWEMPopulation2}
\end{figure}

{{The BNS GW detectabilities for the networks of 2.5th, and 3rd generation GW detectors have been studied in detail in Section \ref{sec:GWResults2}. Based on these results, we now discuss the EM detection probabilities and optimistic EM detection rates for GW-triggered ToO observations.}} 

{{Since the luminosity distributions in different bands are consistent, we only show $g$-band luminosity distributions for the EM signals of detectable GW {{events}} at different GW eras in Figure \ref{fig:GWEMPopulation2}. For the future GW eras of PlusNetwork, LIGO Voyager, and ET\&CE, the critical magnitudes for the detection of EM emissions from all BNS GW events would be $\sim 23.5\,{\rm mag}$, $\sim 25\,{\rm mag}$, and $\gtrsim 26\,{\rm mag}$, respectively. Present and foreseeable future survey projects can hardly find all EM signals of BNS GW events detected during the ET\&CE era. Comparing with the results of adjacent GW eras, one can see there appears little difference in the number of detectable kilonovae {{if adopting a detection depth as the critical magnitude of earlier GW era}}. However, one can find much more remote sGRBs and afterglows in the later GW era. As the search limiting magnitude increases, the amount of detectable kilonovae  would increase exponentially. There is also an exponential increase with a slower rising slope for the amount of afterglows. At the critical magnitude of each era, $\sim80\%$ BNS GW events can observe clear kilonova signals, while afterglows would account for the other $\sim20\%$ BNS GW events. Most of the detectable kilonova-dominated BNS GW events would be not accompanied with the observations of sGRBs. The kilonova events associated with sGRBs like the observations of GW170817/GRB170817A/AT2017gfo would be scarce. }}

\begin{figure*}[htpb]
    \centering
    \includegraphics[width = 0.99\linewidth , trim = 70 160 60 35, clip]{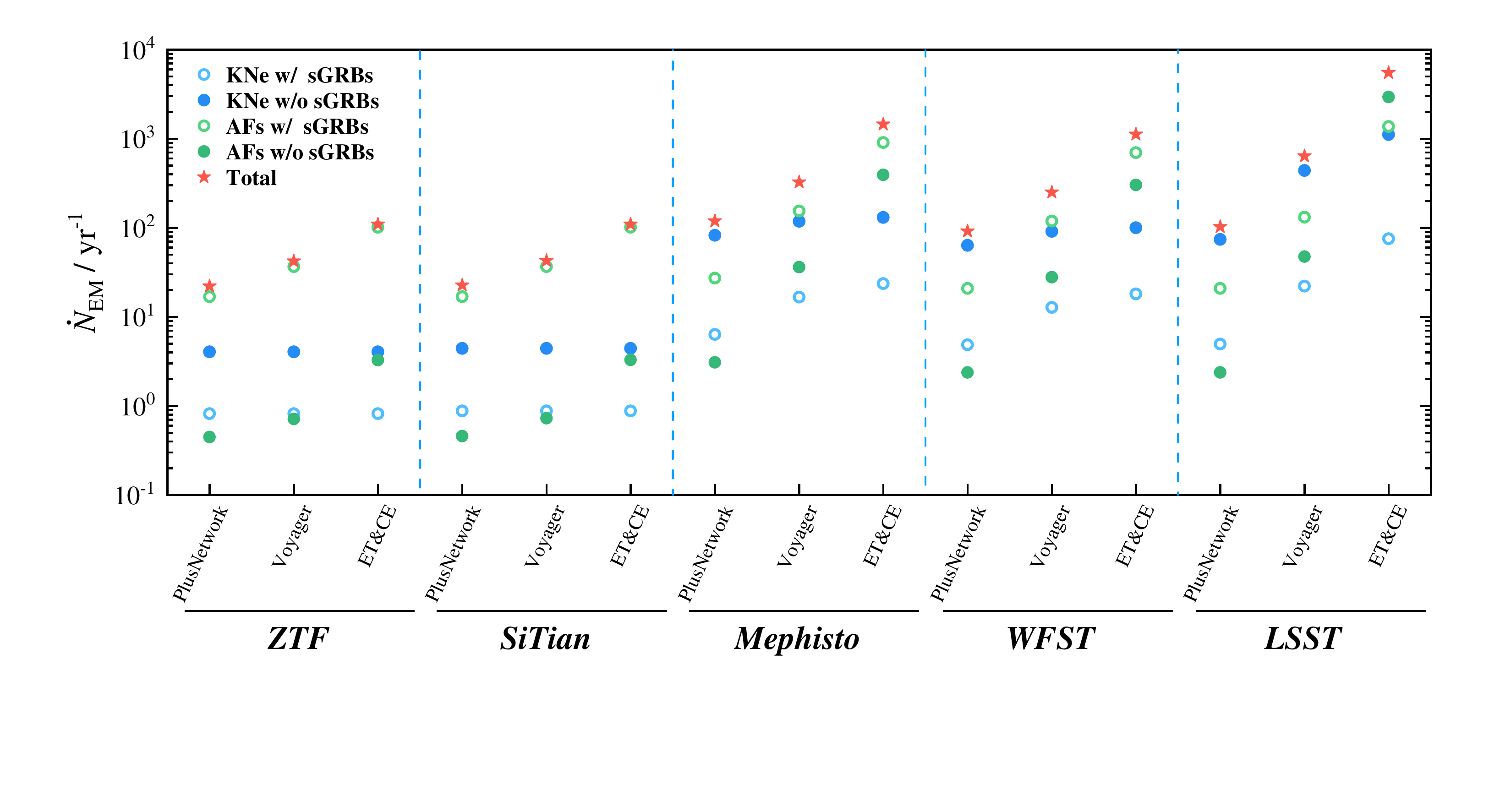}
    \caption{{{$g$-band optimistic detection rates of kilonovae w/ sGRBs (light blue circles), kilonovae w/o sGRBs (dark blue circles), afterglows w/ sGRBs (light green circles), afterglows w/o sGRBs (dark blue circles) and total EM signals (red stars) for specific survey projects (including ZTF, SiTian, Mephisto, WFST, and LSST) in the PlusNetwork, Voyager, and ET\&CE eras.}}}
    \label{fig:ToO}
\end{figure*}

Based on the GW detections in different eras, we then estimate the optimistic EM detection rates for specific survey projects listed in Table \ref{tab:SurveyProject}. In ideal operation conditions of the PlusNetwork and ET\&CE eras, due to precise sky localizations of GW events, these survey projects will be able to cover the sky localizations given by GW detections in a few pointings. Thus, we define $\Omega_{\rm cov} / \Omega_{\rm GW}\sim 1$ in Equation  (\ref{equ:GWEMdetection_rate}) to estimate the EM detection rates. We note that although the current plan at the LIGO Voyager era shows relatively poor sky localization for GW events, we still estimate the EM detection rate of this era by defining $\Omega_{\rm cov}/\Omega_{\rm GW}\sim 1$ {{under the assumption that more GW detectors will}} join the campaign might significantly improve the sky localizations. The optimistic detection rates of EM signals at different GW eras for specific survey projects are displayed in Figure \ref{fig:ToO} and labeled in Table \ref{tab:GWEMDetectablity}. {{By comparing the limiting magnitudes of specific survey projects and the critical magnitudes in the different GW eras (Figure \ref{fig:GWEMPopulation2}), we find that these wide-field surveys (ZTF, SiTian, Mephisto, and WFST) are unlikely to detect a larger number of kilonovae despite the upgraded GW detectors improving BNS detection rates.}} Optimistically, ZTF/SiTian/Mephisto/WFST can detect $\sim5/5/150/120$ kilonovae per year at the 2.5th and 3rd generation era, while $\sim100/300/1200\,{\rm yr}^{-1}$ kilonovae per year can be discovered by LSST during the PlusNetwork/LIGO Voyager/ET\&CE eras, respectively. At later GW era, ToO observations of BNS GW events can always discover more afterglows, almost all of which are associated with sGRBs. {{However, we find a special case that LSST follow-up in the ET\&CE era will detect as much as $\sim75\%$ of detectable afterglows, which will largely orphans unaccompanied by an sGRB.}}

\begin{deluxetable*}{cccccccc}[tpb!]
\tablecaption{{{$g$-band optimistic EM Detection Rates in Each GW Era}}}
\label{tab:GWEMDetectablity}
\tablewidth{0pt}
\tablehead{
\colhead{Sample} &
\colhead{Era} &
\colhead{ZTF} & 
\colhead{SiTian} & 
\colhead{Mephisto} & 
\colhead{WFST} &
\colhead{LSST} 
}
\startdata
\multirow{3}{*}{KNe w/ sGRBs} & PlusNetwork & 0.8 & 0.9 & 6.3 & 4.9 & 4.9 \\
 & Voyager & 0.8 & 0.9 & 16.6 & 12.8 & 22.1 \\
 & ET\&CE & 0.8 & 0.9 & 23.6 & 18.1 & 75.4 \\ \hline
\multirow{3}{*}{KNe w/o sGRBs} & PlusNetwork & 4.0 & 4.4 & 82.4 & 63.4 & 74.0 \\
 & Voyager & 4.0 & 4.4 & 118 & 90.9 & 438 \\
 & ET\&CE & 4.0 & 4.4 & 130 & 100 & $1.11\times10^3$ \\ \hline
\multirow{3}{*}{AFs w/ sGRBs} & PlusNetwork & 16.8 & 16.8 & 27.1 & 20.9 & 20.9 \\
 & Voyager & 36.6 & 36.6 & 154 & 118 & 131 \\
 & ET\&CE & 101 & 101 & 905 & 696 & $1.37\times10^3$ \\ \hline
\multirow{3}{*}{AFs w/o sGRBs} & PlusNetwork & 0.4 & 0.5 & 3.1 & 2.4 & 2.4 \\
 & Voyager & 0.7 & 0.7 & 36.2 & 27.9 & 47.4 \\
 & ET\&CE & 3.3 & 3.3 & 392 & 302 & $2.93\times10^3$ \\ \hline
\multirow{3}{*}{ToTal EM Signals} & PlusNetwork & 22.1 & 22.6 & 119 & 91.5 & 102 \\
 & Voyager & 42.2 & 42.7 & 325 & 250 & 639 \\
 & ET\&CE & 109 & 110 & $1.45\times10^3$ & $1.12\times10^3$ & $5.49\times10^3$
\enddata
\tablecomments{{{The values represent the simulated BNS merger detection rates (in unit of ${\rm yr}^{-1}$) in different GW eras.}}}
\end{deluxetable*}

\section{Conclusions and Discussion \label{sec:conclusion}}

In this paper, based on our model proposed in the companion paper (\citetalias{zhu2021kilonovaafterglow}), we have presented the serendipitous search detectability of time-domain surveys for BNS EM signals, the detectability of GWs for different generations of GW detectors, as well as joint-search GW signals and optical EM counterparts\footnote{GW and EM detectability in the decihertz GW band could be found in \cite{Liu:2022mcd} and \cite{Kang:2022nmz}.}.

{\em Serendipitous observations}\,---\,We have systematically made simulations of optimal search strategy for searching for kilonova and afterglow emissions from BNS mergers by serendipitous observations. For our selected survey projects, which include ZTF/Mephisto/WFST/LSST, we have found that a one-day cadence serendipitous search with an exposure time of $\sim30\,{\rm s}$ can always achieve near maximum detection rates for kilonovae and afterglows. The optimal detection rate of kilonova-dominated (afterglow-dominated) events are $\sim0.3/0.6/1/20\,{\rm yr}^{-1}$ ({$\sim50/60/100/800\,{\rm yr}^{-1}$}), respectively, for the survey projects of ZTF/Mephisto/WFST/LSST. As for the survey array of SiTian project, we have shown that when the array fully operates it will discover more kilonova events if a longer exposure time is adopted. The detection rate of kilonova (afterglow) events could even reach $\sim7(2\times10^3)\,{\rm yr}^{-1}$ by SiTian. The population properties and fading rates of the detectable kilonovae and afterglows have been studied in detail. Our results have shown that afterglows are easier to detect than kilonovae by these survey projects.{{ These afterglows detected via the optically serendipitous observations should be always associated with sGRBs.}} However, present survey projects have not detected as many afterglows as we have predicted. One reason may be that only part of BNS GW events could generate relativistic jets and power bright afterglows \citep[e.g.,][]{sarin2022}. Genuine weather fluctuations and operational issues of optical telescopes might contribute to the deficit of detection. Actual survey observations can hardly always achieve the prospective detection depth and cadence, which could be another cause of the lack of enough afterglow observations. Furthermore, the relatively longer cadence interval for traditional survey projects have been designed to discover ordinary supernovae or tidal disruption events. These cadence intervals are significantly larger than the timescales during which the brightness of afterglow is above the limiting magnitude (as shown in Table \ref{tab:Delta_t}), so afterglow events could be easily missed. Recently, thanks to the improved cadence, ZTF has discovered seven independent optically-discovered GRB afterglows without any detection of an associated kilonova \citep{andreoni2021b}. Among these detected afterglows, at least one event was inferred to be associated with a sGRB. This ZTF observation may support our afterglow simulations, but also show a possible low efficiency of detecting afterglows. Thus, such selection criteria may miss most of kilonova events. Conversely, the low efficiency of afterglow observations also indicate the difficulty for searching for kilonova signals by serendipitous observations. {{\cite{andreoni2021,andreoni2021b} intended to select kilonova and afterglow candidates from survey database by considering recorded sources having rising rates faster than $1\,{\rm mag}\,{\rm day}^{-1}$ and fading rates faster than $0.3\,{\rm mag}\,{\rm day}^{-1}$. When $m_{\rm limit}\lesssim22\,{\rm mag}$, our detailed studies on the population properties of detectable kilonovae and afterglows reveal that their detected fading rates peak at $\sim0-0.1\,{\rm mag}\,{\rm day}^{-1}$ and $\sim1.3\,{\rm mag}\,{\rm day}^{-1}$, respectively.}}

{{{\em GW detections and ToO follow-ups during O4}\,---\,By applying the duty cycle of O3 to simulate the GW observations during O4, we predict that one can detect $\sim11\,{\rm yr}^{-1}$ BNS GW events with a median detectable distance at $z\sim0.040$ and a horizon at $z_{\rm max}\sim0.084$. The median sky localization area is expected to be $\sim10\,{\rm deg}^2$ for detectable BNS GW events in O4. Based on the public alert distributions in O3, \cite{petrov2022} suggested that the threshold S/N for the detection of BNSs might be lower (i.e., S/N$>9$). Following their suggestions, the detection rate of BNS mergers would be higher and the median GW sky localization area would be larger in O4 by comparison with our simulations. In this paper, we adopt their respective design sensitivities to simulate GW detections and ToO follow-ups of GW triggers since their sensitivities are dynamic and change over time. After the completion of this work, we notice that detection sensitivities of H1, L1, V1, and K1 in upcoming O4 have been updated\footnote{\url{https://observing.docs.ligo.org/plan/}.}. Based on their latest detection sensitivities, our simulations for the detection rate, detectable distance, and sky localization might be slightly better than they will really be. During O4, our simulations show that ZTF/Mephisto/WFST/LSST will detect $\sim5/4/3/3$ kilonovae ($\sim1/1/1/1$ afterglows) per year, respectively. Most of these detectable afterglows are expected to be associated with sGRBs, while only $\lesssim5\%$ kilonovae can simultaneously detect their associated sGRBs after GW triggers. }}

{{{\em GW detections and ToO follow-ups at the 2.5th and 3rd generation eras}\,---\,We have carried out detailed calculations of the detection capabilities of the 2.5th and 3rd generation detector networks in the near future for BNS GW signals. Optimistically, we show that the GW detection rate and detection horizon for the PlusNetwork are $\sim210\,{\rm yr}^{-1}$ and $z_{\rm max}\sim0.2$, respectively. Most of detectable BNS mergers will be localized to $\lesssim10\,{\rm deg}^2$, which are always smaller than the FoV of most of the survey projects. For the LIGO Voyager in the 3rd generation era, the optimal detection rate can be increased to $\sim1,800\,{\rm yr}^{-1}$ and the detection distance would be twice compared with the last era, i.e., a horizon of $z_{\rm max}\sim0.4$. The ET\&CE network is expected to detect all BNS merger events in the entire universe, with detection rates $\sim2.4\times10^5\,{\rm yr}^{-1}$. As the sensitivity of GW detectors increases, BNS events at high redshifts gradually dominate the detected events. At this era, the detection rate is mainly dominated by BNS mergers at $z \sim 1$. The median locaization for BNS mergers at $z\sim0.5$ ($z\sim1$) is shown to be $\sim1\,{\rm deg}^2$ ($\sim10\,{\rm deg}^2$). In the PlusNetwork and LIGO Voyager eras, the critical magnitudes for the detection of EM emissions from all BNS GW events would be $\sim 23.5\,{\rm mag}$ and $\sim 25\,{\rm mag}$, respectively. At the critical magnitude of each era, $\sim80\%$ BNS GW events can observe clear kilonova signals, while afterglows would account for the other $\sim20\%$ BNS GW events. ZTF/SiTian/Mephisto/WFST can optimistically detect $\sim5/5/150/120$ kilonovae per year at the 2.5th and 3rd generation era, while $\sim100/300/1200\,{\rm yr}^{-1}$ kilonovae per year can be discovered by LSST during the PlusNetwork/LIGO Voyager/ET\&CE eras, respectively. At later GW era, ToO observations of BNS GW events can always discover more afterglows, almost all of which are associated with sGRBs. Present and foreseeable future survey projects can hardly find all EM signals of BNS GW events detected during the ET\&CE era.}} By assuming a single-Gaussian structured jet model \citep[e.g.,][]{zhang2002}, we have shown that GW170817-like events, which can be simultaneously observed as an off-axis sGRB and a clear kilonova, may be scarce. In order to explain the sGRB signal of GW170817/GRB170817A, a two-Gaussian structured jet model may be required \citep{tan2020}. Future multi-messenger detection rates of sGRBs, kilonovae and afterglows can be used for constraining the jet structure.

In this paper, we adopt an AT2017gfo-like model as our standard kilonova model to calculate the kilonova detectability of serendipitous and GW-triggered ToO observations. However, many theoretical works in the literature \citep[e.g.,][]{kasen2013,kasen2017,kawaguchi2020,kawaguchi2021,darbha2020,korobkin2021,wollaeger2021} show that BNS kilonova should be diverse which may depend on the mass ratio of binary and the nature of the merger remnant. The possible energy injection from the merger remnant, e.g., due to spindown of a post-merger magnetar \citep{yu2013,yu2018,metzger2014,ai2018,li2018,ren2019}\footnote{The dissipation of wind from remnant magnetar \citep{zhang2013} or interaction between the relativistic magnetar-driven ejecta and the circumstellar medium \citep{gao2013,liu2020} may also produce additional optical emission.} or fall-back accretion onto the post-merger BH \citep{rosswog2007,ma2018} could significantly increase the brightness of the kilonova. The diversity of kilonova and potential energy injection may affect on the final detection rate of kilonova, which will be studied in future work.

\software{\texttt{POSSIS} \citep{bulla2019,coughlin2020measuring}; Matlab, \url{https://www.mathworks.com}; Python, \url{https://www.python.org}; LALSuite, \citep{lalsuite}}

\acknowledgments
{{We thank the anonymous {{referees}} for constructive comments.}}
We thank Xue-Feng Wu and Jiming Yu for valuable comments. The work of J.P.Z is partially supported by the National Science Foundation of China under Grant No. 11721303 and the National Basic Research Program of China under grant No. 2014CB845800. Y.P.Y is supported by National Natural Science Foundation of China grant No. 12003028, the National Key Research and Development Program of China (2022SKA0130101), and the China Manned Spaced Project (CMS-CSST-2021-B11). Z.J.C is supported by the National Natural Science Foundation of China (No. 11690023). H.G. is supported by the National Natural Science Foundation of China under Grant No. 11690024, 12021003, 11633001. Y.W.Y is supported by the National Natural Science Foundation of China under Grant No. 11822302, 11833003. L.S. is supported by the National Natural Science Foundation of China under Grant No. 11975027, and the Max Planck Partner Group Program funded by the Max Planck Society.

\appendix

\section{Amplitude Spectral Density} \label{app:ASD}

\begin{figure*}[h]
   \centering
    \includegraphics[width = 0.32\linewidth , trim = 45 30 88 60, clip]{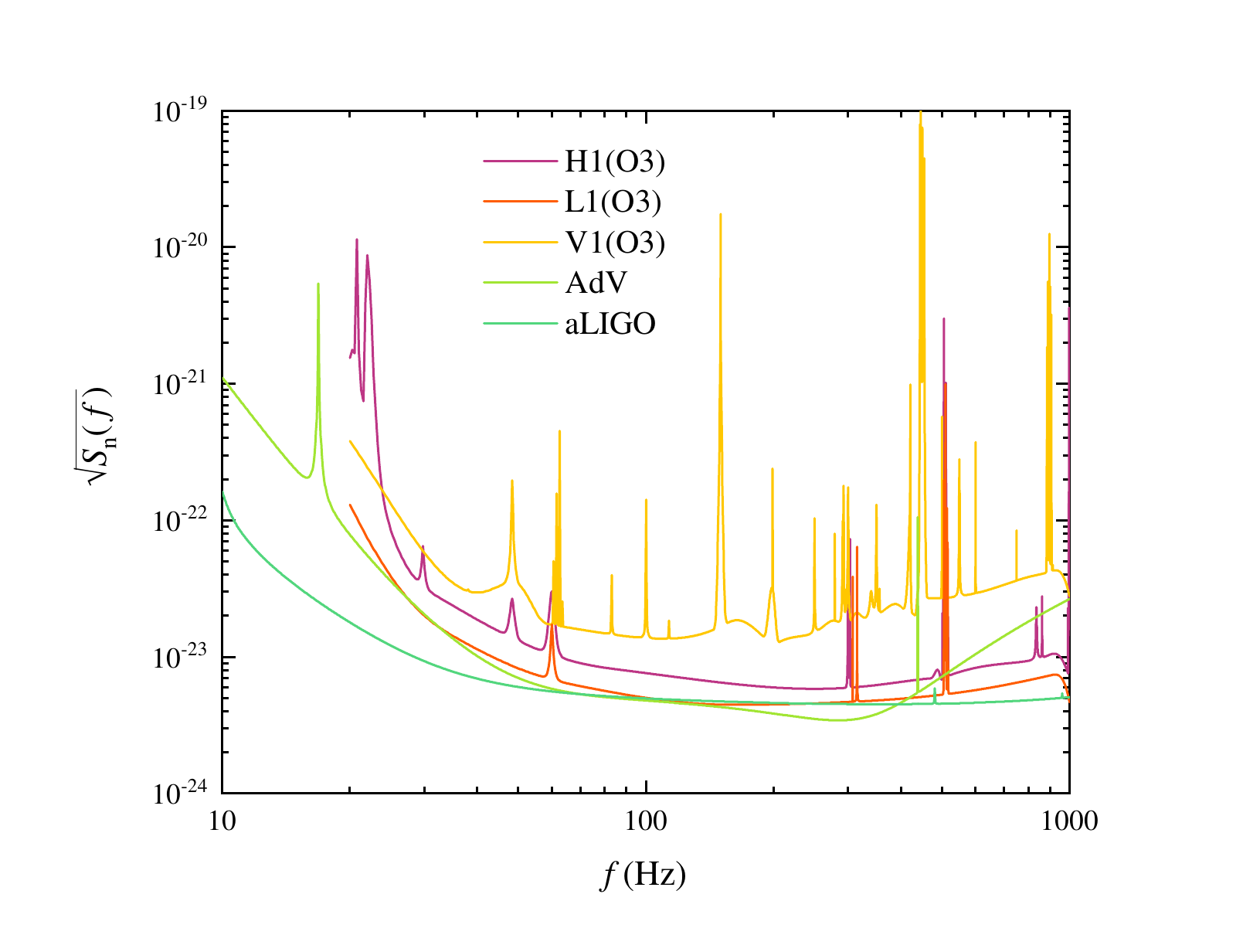}
    \includegraphics[width = 0.32\linewidth , trim = 45 30 88 60, clip]{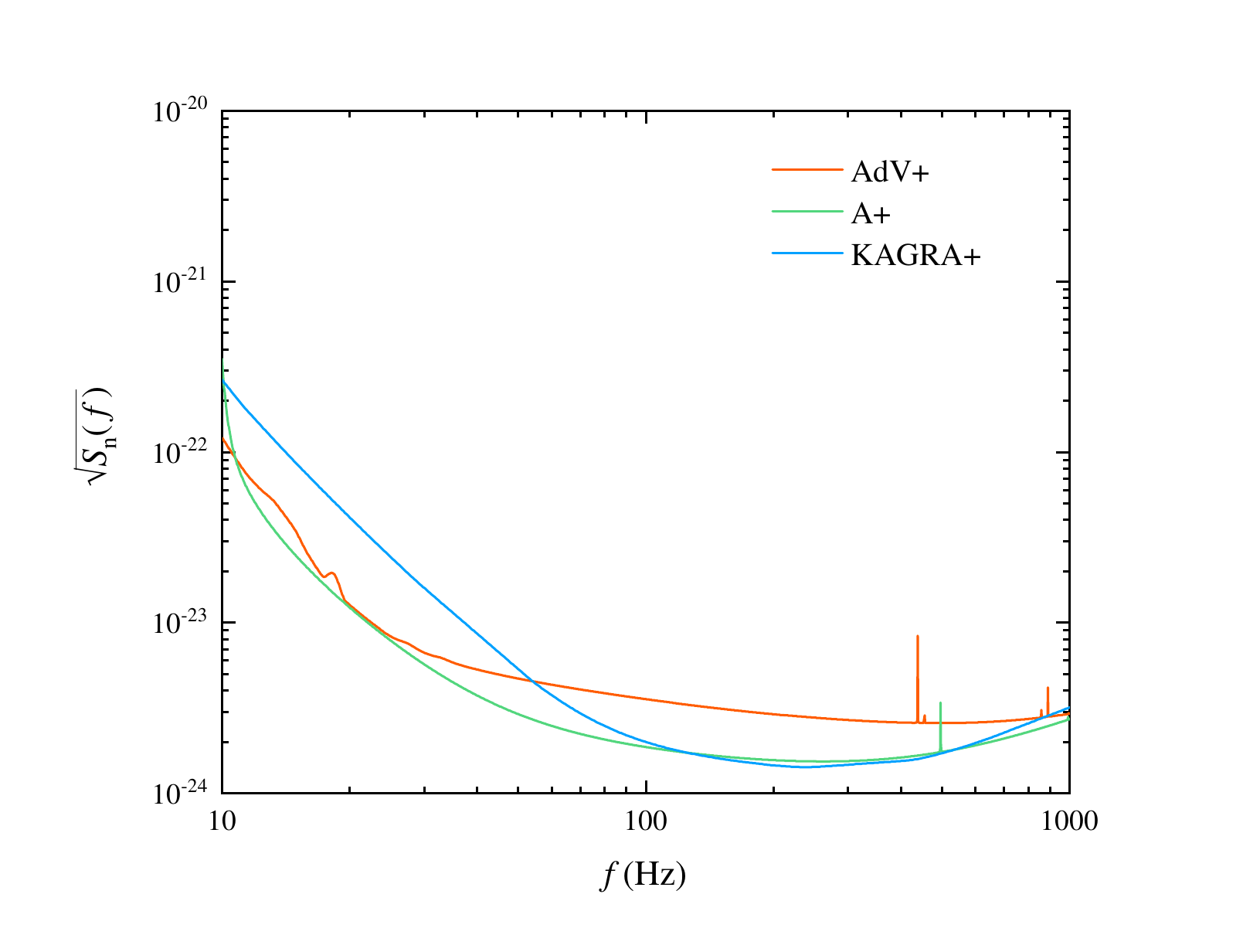}
    \includegraphics[width = 0.32\linewidth , trim = 45 30 88 60, clip]{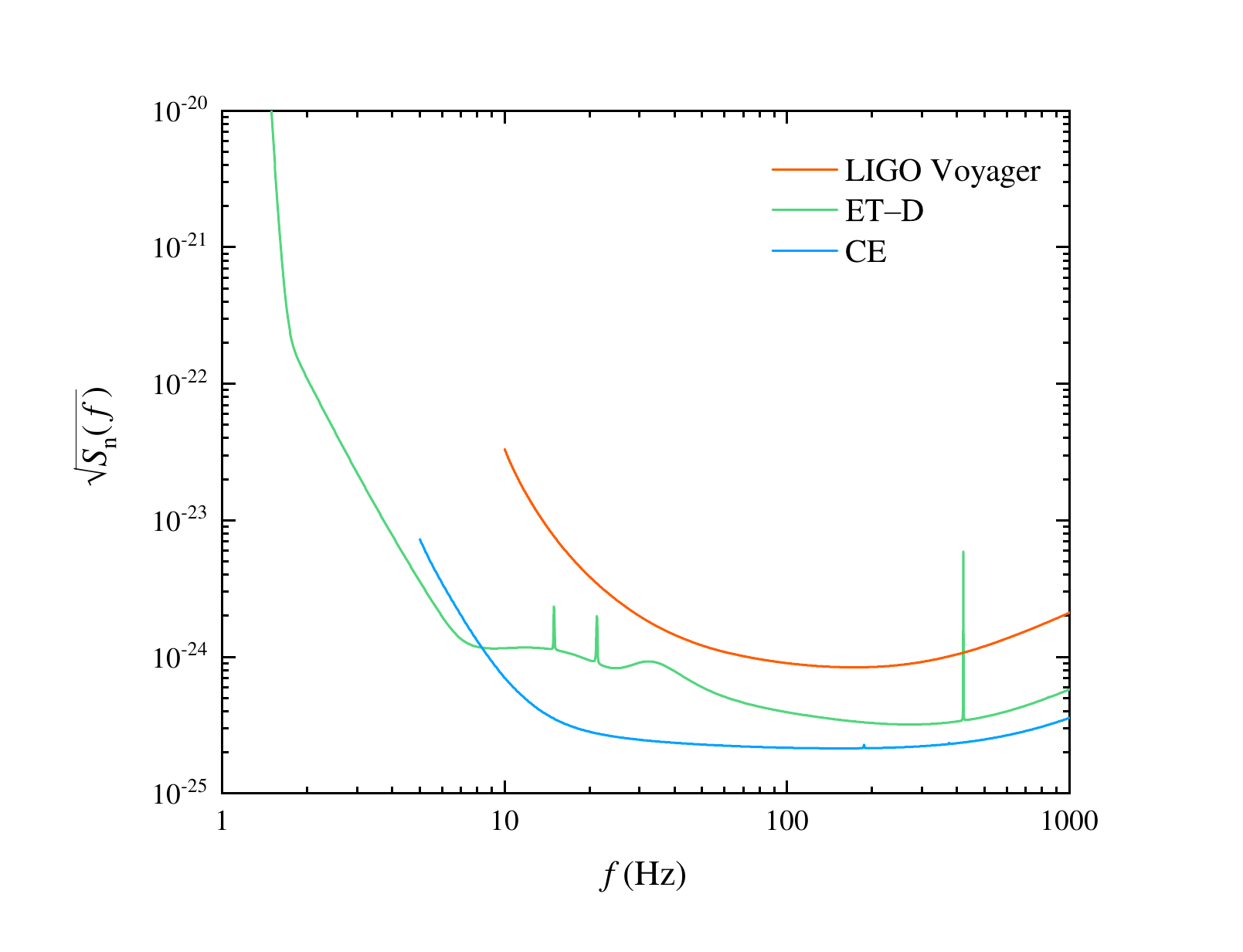}
    \caption{Left panel: the design sensitivity curves of 2nd generation GW detectors and O3 sensitivity curves. Middle panel: the design sensitivity curves of 2.5th generation GW detectors. Right panel: the design sensitivity curves of 3rd generation GW detectors.}
    \label{fig:ASD}
\end{figure*}

The ASD sensitivity curves of GW detectors used in our calculations are presented in Figure \ref{fig:ASD}. For O3, we adopt the GW190814's sensitivity curves\footnote{\url{https://dcc.ligo.org/P2000183/public}} as the O3 sensitivity. The detector sensitivities during HLV (O4), PlusNework and LIGO Voyager era are adopted from the public data\footnote{\url{https://dcc.ligo.org/LIGO-P1200087-v42/public}}\footnote{\url{https://dcc.ligo.org/LIGO-T1500293/public}}. The sensitivity curves of ET and CE used in this paper come from the official websites\footnote{\url{http://www.et-gw.eu/index.php/etsensitivities}}\footnote{\url{https://dcc.cosmicexplorer.org/cgi-bin/DocDB/ShowDocument?docid=T2000017}}. 

\begin{figure}[h]
   \centering
    \includegraphics[width = 0.5\linewidth , trim = 45 30 88 60, clip]{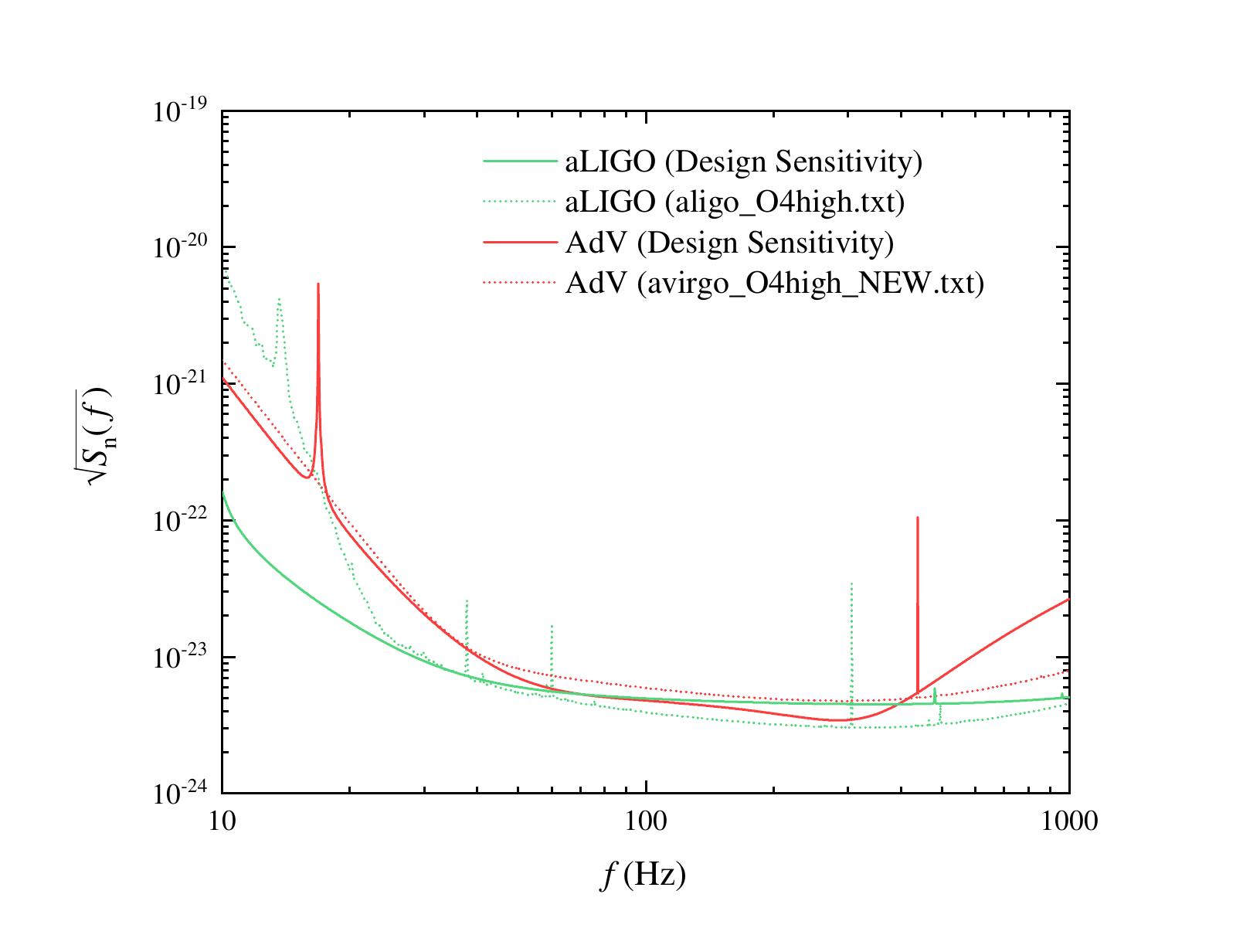}
    \caption{{{Design sensitivity curves (sold lines) and latest O4 sensitivity curves (dashed lines) of Advanced LIGO (red lines) and Advanced Virgo (green lines).}}}
    \label{fig:ASD_Comparsion}
\end{figure}

{{For 2nd generation detectors, we also compare their design sensitivity curves to the latest sensitivity curves released on April 6th, 2022\footnote{\url{https://dcc.ligo.org/LIGO-T2000012-v2/public}} in Figure \ref{fig:ASD_Comparsion}. }}

\bibliography{ms}{}
\bibliographystyle{aasjournal}

\end{document}